%% file: diss.tex
\documentclass[12pt,oneside]{report}
\usepackage{puthesis}
\renewcommand{\baselinestretch}{1.7}

\begin{document}
\input{defs}
\pagenumbering{roman}
\pagestyle{plain} 
\include{title}
\addcontentsline{toc}{chapter}{Abstract}
\include{abstract}

\include{quotes}

\tableofcontents     \vfill\cleardoublepage\thispagestyle{empty}
\addcontentsline{toc}{chapter}{Acknowledgments}
\thispagestyle{plain}

\include{thanks}        \vfill\cleardoublepage\thispagestyle{empty}
\addcontentsline{toc}{chapter}{Notation}
\thispagestyle{plain}
\include{notation}        \vfill\cleardoublepage\thispagestyle{empty}
\renewcommand{\baselinestretch}{1.7}
\vfill\cleardoublepage\thispagestyle{empty}
\pagestyle{headings} \pagenumbering{arabic}
\chapter{Black Holes From The Outside}
\label{ch:intro}
\input{intro}

\vfill\cleardoublepage\thispagestyle{empty}
\chapter{Classical Theory: The Membrane Paradigm}
\label{ch:membrane}
\input{membrane}
\vfill\cleardoublepage\thispagestyle{empty}
\chapter{Semi-Classical Theory: Thermodynamics}
\label{ch:instanton}
\input{instanton}

\vfill\cleardoublepage\thispagestyle{empty}
\chapter{The Causal Structure Of Evaporating Black Holes}
\label{ch:vaidya}
\input{vaidya} 
\vfill\cleardoublepage\thispagestyle{empty}
\addcontentsline{toc}{chapter}{References}
\bibliographystyle{phd}
\bibliography{thesis}
\end{document}

%% file: defs.tex
\def \d {\delta}
\def \pl {\partial}
\def \cc {\overline}
\def \f {\frac}
\def \rg {\sqrt {-g} \,}
\def \rh {\sqrt {-h} \,}
\def \lf {\left (}
\def \rt {\right )}
\def \th {\theta}
\def \sg {\sigma}
\def \gm {\gamma}
\def \w {\omega}
\def \eps {\epsilon}
\def \del {\nabla}
\def \lm {\lambda}
\def \gm {\gamma}
\def \l2 {\lambda _2}
\def \rg {\sqrt {-g} \,}
\def \rh {\sqrt {-h} \,}
\def \mn {\mu \nu}
\def \phi {\varphi}
\def \wps {{\omega \psi \over \pi}}
\def \Mw2 {\lf M - \f{\scriptstyle \w}{2} \rt}
\def \root {\sqrt{M^2 - Q^2}}
\def \Root {\sqrt{2Mr - Q^2}}
\def \tu {\tilde u}
\def \vaidya {\lf 1 - {2M(u) \over r} + {Q^2(u) \over r^2} \rt}

%% file: title.tex
\thispagestyle{empty}

\begin{center}

\vspace*{0.6in}

MEMBRANE HORIZONS: THE BLACK HOLE'S NEW CLOTHES

\vspace*{1.2in}

Maulik Kirit Parikh \\

\vspace*{1.2in}

A DISSERTATION \\
PRESENTED TO THE FACULTY \\
OF PRINCETON UNIVERSITY \\
IN CANDIDACY FOR THE DEGREE \\
OF DOCTOR OF PHILOSOPHY \\

\vspace*{0.6in}

RECOMMENDED FOR ACCEPTANCE \\
BY THE DEPARTMENT OF \\
PHYSICS \\

\vspace*{0.6in}

October 1998

\end{center}

%% file: abstract.tex
\centerline{\bf Abstract}

\vspace*{0.20in}

This thesis addresses some classical and semi-classical aspects of black holes,
using an effective membrane representation of the event horizon.
This ``membrane paradigm'' is the remarkable view that, 
to an external observer, a black hole appears to behave exactly like a
dynamical fluid membrane, obeying such pre-relativistic equations as Ohm's
law and the Navier-Stokes equation. It has traditionally been derived 
by manipulating the equations of motion. Here, however, the equations are
derived from an underlying action formulation which has the advantage of
clarifying the paradigm and simplifying the derivations, in addition to
providing a bridge to thermodynamics and quantum mechanics. Within this 
framework, previous membrane results are derived and extended to dyonic 
black hole solutions. It is explained how an action can produce dissipative 
equations. The classical portion of the study ends with a demonstration of
the validity of a minimum entropy production principle for black holes.

Turning next to semi-classical theory, it is shown that familiar 
thermodynamic properties of black holes also emerge from the membrane 
action, via a Euclidean path integral. In particular, the membrane
action can account for the hole's Bekenstein-Hawking entropy, including
the numerical factor. Two short and direct derivations of Hawking radiation 
as an instanton process are then presented. The first is a tunneling 
calculation based on particles in a dynamical geometry, closely
analogous to Schwinger pair production in an electric field.
The second derivation makes use of the membrane representation of the 
horizon. In either approach, the imaginary part of the action 
for the classically forbidden process is related to the Boltzmann
factor for emission at the Hawking temperature. But because
these derivations respect conservation laws, the exact result contains a
qualitatively significant correction to Hawking's thermal spectrum. 

Finally, by extending the charged Vaidya metric to cover all of spacetime,
a Penrose diagram for the formation and evaporation of a charged black hole
is obtained. It is found that the spacetime following the evaporation of
a black hole is predictable from initial conditions, provided
that the dynamics of the time-like singularity can be calculated.

%% file: quotes.tex
\thispagestyle{empty}

\clearpage

\vspace*{1.5in}

``Coffee ... is the fuel of science'' 

\hspace*{1.5in} -- John Archibald Wheeler, overheard at tea

%% file: thanks.tex
\centerline{\bf Acknowledgments}

\vspace*{0.20in}

First, I owe immeasurable thanks to Frank Wilczek, my supervisor, 
who guided this thesis from beginning to end. He not only treated me to
a perfect blend of advice, encouragement, ideas, and independence,
he also taught me a style of physics.

In addition, I have benefited from interactions with the physics 
community. Other physicists on whom I have imposed in various ways include 
David Gross, Igor Klebanov, P. J. E. Peebles, Alexandre Polyakov, 
Joseph Taylor, Kip Thorne, and Herman Verlinde.

The foundations of much of my understanding lay in structured study groups 
and impromptu discussions. For these I would like to thank my 
physics friends, fourth-floor fixtures John Brodie, Seth Bruder, 
Ekaterina Chashechkina, Erick Guerra, Marco Moriconi, 
Christopher Wiggins, and H. El\c{c}in Yildirim.

More generally, I must thank my circle of friends and relatives, 
a crucial supporting cast. It was by their continual presence, by
their admonitions-blandishments-cajolery, that this thesis did not 
take considerably longer. In particular, my deepest gratitude 
goes to Otute Akiti, Katerina Papadaki, Nina and Indra Shah, Wai-Hong Tham,
and Yingrui Yang.

And, finally, I owe thanks to my sister, Anokhi, and to Jyoti and Kirit, 
my parents -- to whom this thesis is dedicated -- 
for their love and nonstop support.

To all the above, and to any I may have missed, thank you.

%% file: notation.tex
\centerline{\bf Notation}

\vspace*{0.20in}

In this work, we use lowercase indices for four-dimensional tensors
and uppercase indices for the two-dimensional tensors that occupy
space-like sections of the horizon. Repeated indices are implicitly
summed over.

The spacetime metric tensor is denoted by $g_{ab}$, the induced
metric on the time-like stretched horizon is $h_{ab}$, the metric on a
space-like slice of spacetime is ${}^3g_{ab}$, and the metric
on a space-like section of the horizon is written $\gamma _{AB}$. 
Correspondingly, we denote the 4-covariant derivative by $\del_a$, the 
3-covariant derivative on the stretched horizon by $|a$, and the 
2-covariant derivative on a space-like slicing by $\|A$.

We take the spacetime metric to have signature $(-+++)$. Our 
sign conventions are those of Misner, Thorne, and Wheeler, 
with the exception of the extrinsic curvature which we define to have
a positive trace for a convex surface.

Throughout, we use geometrized units in which Newton's constant, $G$, and the
speed of light, $c$, are set to one. We shall also usually set Planck's
constant, $\hbar$, and Boltzmann's constant, $k_B$, to one. The appropriate
factors of these constants may always be restored through dimensional analysis.

%% file: intro.tex
\section{Causality and the Horizon}

A black hole is a region of spacetime from which, crudely speaking, there
is no escape. The boundary of a black hole is called an
{\it event horizon} because an outside observer is unable to observe events on 
the other side of it; light rays from inside cannot propagate out, hence 
``black'' hole.

{}From these basic definitions, several properties of the horizon 
emerge as consequences. We emphasize four that will be especially
important to this study:

\begin{quote}
$\bullet$ The horizon is a {\it causal} boundary. Since no signal can get out,
the inside of a black hole cannot influence the outside; we say
that the black hole's interior is causally disconnected from the outside.
\end{quote}

\begin{quote}
$\bullet$ The spacetime containing a solitary black hole is inherently 
{\it time-reversal asymmetric}. The time reverse of infall -- escape -- 
does not occur, so a direction for the arrow of time is implicit.
\end{quote}

\begin{quote}
$\bullet$ 
Because nothing can travel faster than light, the event horizon must be
a {\it null hypersurface}, a surface along which light travels.
\end{quote}

\begin{quote}
$\bullet$ The definitions are really {\it global} statements, 
requiring knowledge of extended regions of spacetime. 
For, to know whether one is inside a black hole, one has to to know whether
one can eventually escape, and that requires knowing one's future.
Thus it is impossible to divine on the basis of local -- here and now --
measurements whether one is already inside a black hole. Indeed, the
event horizon is an unmarked border, with no local signifiers of 
its presence such as a divergent curvature scalar.
In fact, in nonstationary situations, an event horizon 
may be present even in flat space.
\end{quote}

Now, since the inside of a black hole is causally disconnected from the
outside, classical physics for an outside observer should be independent
of the black hole's interior. One can therefore ask what an
observer who always stays outside the horizon sees.
As a number of authors have discovered, the answer is remarkable.
The same horizon which we have just noted is invisible to an 
infalling observer, appears, to an outside observer, transformed into
a dynamical membrane with tangible, {\it local} physical properties such as 
resistivity and viscosity. Moreover, the equations of motion governing
the membrane are {\it nonrelativistic} -- Ohm's law and the 
Navier-Stokes equation -- even though they 
describe what is a quintessentially relativistic
entity. Furthermore, these equations are dissipative, even though they
may be derived in an action formulation. And the 
analogies with membranes go quite far; the horizon shares not only the
classical properties of real membranes, but also their semi-classical ones.
In particular, a membrane description of the horizon is able to account
for such {\it thermodynamic} notions as entropy and the black hole's tendency
to radiate as if it possessed a temperature. The main difference between
horizon membranes and real membranes such as soap bubbles is
that the black hole membrane is {\it acausal}, reacting to events before
they occur. And, of course, it is not really there; as we have already noted,
the observer who falls through the event horizon sees nothing, nothing at all.

That elaborate illusion is the main subject of this dissertation. 

\section{Background and Overview}

The earliest relativistic studies of spherically symmetric 
gravitational sources began with the Schwarzschild line element,
\begin{equation}
ds^2 = - \lf 1 - {2M \over r} \rt \, dt^2 + {dr^2 \over 1 - {2 M \over r}} + 
r^2 \lf d \theta ^2 + \sin ^2 \theta d \phi ^2 \rt \; ,
\end{equation}
from which it seemed that there was a singularity at the Schwarzschild
radius, $r = 2M$. It was not realized initially that this singularity was 
merely a {\it coordinate singularity}, an artifact of using a pathological 
system of coordinates. Instead, it was
thought that the region $r \leq 2M$ was somehow unphysical. Einstein
himself contributed to this faulty view by showing that no stationary
configuration of matter could exist inside the Schwarzschild radius,
in the process repeating his cosmological ``blunder'' of overlooking 
nonstationary configurations (for a history, see \cite{israelhistory}).

But later, with the work of Oppenheimer and Snyder, it became clear that a 
collapsing star could actually pass through the Schwarzschild radius 
and would effectively disappear from the outside, becoming a ``black hole''. 
Thus, Wheeler:
\begin{quote}
... it becomes dimmer millisecond by millisecond, and in less than a second is
too dark to see. What was once the core of a star is no longer visible.
The core like the Cheshire cat fades from view. One leaves behind only
its grin, the other, only its gravitational attraction ...
\end{quote}
The idea that one could fall through the Schwarzschild radius opened up 
the study of the inside of black holes and of its relationship to the
outside, a chain of work which culminated in the modern view of the 
event horizon as a causal boundary with the properties described above.
Finally, in the work of Damour, Thorne, and others, the
dynamical nature of the boundary, and its membrane interpretation,
was clarified.

Our own study begins in Chapter 2 by considering these classical aspects of 
the event horizon. Actually, because null surfaces have a number of degenerate 
properties, mathematical descriptions of the event horizon 
are a little inconvenient. However, if, instead of the
light-like event horizon, we consider a {\it time-like} surface infinitesimally
outside it, then we can gain mathematical ease while still retaining 
the event horizon's other properties. This time-like surrogate, because 
it encloses the event horizon, is known as the {\it stretched horizon}, 
and it is this surface that we will be working with mostly. 
After collecting some mathematical facts about the stretched horizon, 
we demonstrate that its equations of motion can be derived from an 
action principle \cite{membraneaction}, and that they can be considered to
describe a dynamical membrane. We provide concrete
examples for the membrane interacting with electromagnetic, gravitational,
and axidilaton fields, including derivations of Ohm's law, the Joule
heating law, and the Navier-Stokes equation as they apply to black holes.
This is followed by a discussion of the origin of dissipation, and its
relation to the breaking of time-reversal symmetry. We conclude the
chapter by providing a Hamiltonian formulation, and showing that, for
quasi-stationary black holes, the equations of motion follow also from a
minimum heat production principle, as advocated by Prigogine.

When the fields the black hole interacts with are not classical but quantum
fields, new phenomena emerge. Because the part of the wave function that
lies within the black hole is inaccessible to the outside observer, 
we expect the black hole to have an entropy, associated with ignorance of
those degrees of freedom. In addition, particles just inside the 
black hole can now escape, since the quantum uncertainty principle 
blurs the exact position of the horizon or, equivalently, of the particles.
Thus black holes radiate \cite{swh}. Moreover, the spectrum of the
outgoing radiation is, at least to first approximation, the Planckian
spectrum of a thermal black-body. Hence one can associate
a {\it temperature} to a black hole.

Historically, the concept of temperature and entropy for black holes was
foreshadowed by several results in the classical theory 
\cite{christo,christoruff,fourlaws,arealaw}, in which something
proportional to the surface gravity played the role of temperature, and 
some factor times the horizon's surface area mimicked the entropy.
As we shall see the classical membrane equations similarly hint at
thermodynamics. However, to go beyond analogies to actually identify the
corresponding quantities requires quantum mechanics, and in particular a
mechanism by which black holes can radiate.

Of course, escape from a black hole contradicts its very definition.
It is easy to see that, when pushed to its logical limit, 
the whole concept of a black hole becomes unreliable, an essentially
classical description. Thus, although quantum field theory around 
black holes does resolve some thermodynamic
paradoxes, the tension between quantum mechanics and general relativity 
remains. In particular, a new {\it information puzzle} arises if the 
prediction of thermal radiation is taken literally. Specifically, 
the question arises whether the initial state that formed
the black hole can be reconstructed from the outgoing radiation.
The main problem here is this: when matter falls into a black hole, the
external configuration is determined by only a few quantities -- this is
the content of the ``no-hair theorem'' \cite{nohair}. At this point, one could
consider the information to reside inside the black hole. However, once
the black hole starts to radiate, the inside disappears and one is 
essentially left with the outgoing radiation. But now purely thermal 
radiation is uncorrelated, so the outgoing radiation does not
carry sufficient information to describe what made up the black hole.

If this is true, if the radiation is purely thermal as has been claimed
\cite{infoloss}, then physics loses its 
predictive power since the final state is not uniquely determined by 
the initial state, with the many different initial configurations of matter
that could form a black hole all ending up as the same bath of thermal
radiation. At stake in the information puzzle is the very idea that 
the past and the future are uniquely connected.

To decisively settle such
questions, one really needs a quantum theory of gravity, one in which the
quantum states that are counted by the black hole's entropy can be
precisely enumerated, and any nonthermal aspects of black hole 
evaporation can be probed through scattering calculations. 
However, string theory, currently our only candidate theory of quantum 
gravity, is at present limited in its ability to answer some important 
questions about spacetime. Hence, in Chapter 3, we take an alternative 
approach: we truncate Einstein gravity at a semi-classical level, with the 
understanding that the infinite higher order corrections will 
actually be rendered finite by a suitable theory of quantum gravity. 
Within this approximation, we find continued success for the membrane 
action. By means of a Euclidean
path integral, we find that the membrane action is also responsible for
the black hole's Bekenstein-Hawking entropy, including the important
numerical factor.

Next, we turn to black hole radiance. In previous derivations of 
Hawking radiation, the origin of the radiation has been somewhat obscure. 
Here we provide two short and physically appealing semi-classical 
derivations of Hawking radiation \cite{instanton}. 
First, we show that the heuristic
notion of black hole radiance as pair production through tunneling does
indeed have quantitative support. The imaginary part of the action 
for tunneling across the horizon is related to the emission rate, just
as in Schwinger pair creation in an electric field. Alternatively, an outside
observer can consider the outgoing flux as consisting of spontaneous emissions
from the membrane, rather than as particles that have tunneled across the
horizon. Since we already have an action for the membrane, we can compute the 
rate for the membrane to shrink spontaneously; 
the emission rate agrees with the tunneling calculation.

We find that the probability for emission is approximately consistent
with black-body emission. However, unlike the original derivations of black
hole radiance, our calculations respect energy conservation. Indeed, 
energy conservation is a fundamental requirement in tunneling, one that
drives the dynamics, and without which our calculations do not even go 
through. The constraint that energy
be conserved modifies the emission rate so that it is not {\it exactly} 
Planckian; there is a nonthermal correction to the spectrum.
The correction to the Hawking formula is small when the outgoing
particle carries away only a small fraction of the black hole's mass.
However, it is qualitatively significant because nonthermality
automatically implies the existence of correlations in the outgoing 
radiation, which means that at least some information must be returned.

Energy conservation has another, long-term, consequence: the black hole
can actually disappear entirely if all its mass and charge are radiated 
away. The question of the causal structure of a spacetime containing 
an evaporating black hole is of some interest, not least because 
it is closely related to the possibility of information retrieval. 
In Chapter 4, we examine the causal structure of such a spacetime.
Approximating the radiating black hole geometry by the 
Vaidya solution that describes the spacetime outside a radiating star, 
we construct a spacetime picture of the formation and subsequent 
evaporation of a charged black hole \cite{vaidyaglobal} in a special case.
We find that the resultant Penrose diagram is predictable in
the sense that post-evaporation conditions are causally dependent on
initial conditions, a result consistent with information conservation.

%% file: membrane.tex
\section{Introduction}         
The event horizon of a black hole is a peculiar object: it is a mathematically
defined, locally undetectable boundary, a surface-of-no-return 
inside which light cones tip over and
``time'' becomes spatial (for a review see, e.g., \cite{he}).
Otherwise natural descriptions of physics often have trouble 
accommodating the horizon; as the most primitive example, the 
familiar Schwarzschild metric has a coordinate singularity there. 
Theories of fields that extend to the horizon face the additional challenge of
having to define boundary conditions on a surface that is infinitely
red-shifted, has a singular Jacobian, and possesses a normal vector which is
also tangential. These considerations might induce one to believe that
black hole horizons are fundamentally different from other physical entities.

On the other hand, further work has established a 
great variety of analogies between the
horizon and more familiar, pre-relativistic bodies. In addition to
the famous four laws of black hole thermodynamics 
\cite{christo,christoruff,bek,fourlaws},
which are global statements, there is also a precise local mechanical and
electrodynamic correspondence. In effect, it has been shown 
\cite{hanniruff,damour78,znajek,damourphd,damour82} 
that an observer who remains outside a black hole perceives the
horizon to behave according to equations that describe a fluid bubble with
electrical conductivity as well as shear and bulk viscosities.
Moreover, it is possible to define a set of local surface densities, such as
charge or energy-momentum, which
inhabit the bubble surface and which obey conservation laws.
Quite remarkably, a general-relativistically
{\em exact} calculation then leads, for arbitrary nonequilibrium
black holes, to equations for the horizon which can be
precisely identified with Ohm's law,
the Joule heating law, and the Navier-Stokes equation.

These relations were originally derived for the mathematical, or true, event
horizon. For astrophysical applications it became more convenient to consider
instead a ``stretched horizon,'' a (2+1)-dimensional time-like surface located 
slightly outside the true horizon.
Because it has a nonsingular induced metric, the stretched horizon
provides a more tractable boundary on which to anchor external fields; 
outside a complicated boundary layer, the equations governing the 
stretched horizon are to an excellent approximation \cite{tmac,pricethorne}
the same as those for the true horizon. This view of
a black hole as a dynamical time-like surface, or membrane, has been called
the membrane paradigm \cite{mempar}.

Most of the mentioned results have been derived through general-relativistic
calculations based on various intuitive physical arguments. In this chapter,
we show that the gravitational and electromagnetic descriptions of the
membrane can be derived systematically, directly, and more simply 
from the Einstein-Hilbert or Maxwell actions. Aside from the appeal 
inherent in a least action
principle, an action formulation is a unifying framework which is easily
generalizable and has the advantage of providing a bridge to thermodynamics
and quantum mechanics (see \cite{carlip} for related work).

The key idea in what follows is that, since
(classically) nothing can emerge from a black hole, 
an observer who remains outside a black hole cannot be affected by the 
dynamics inside the hole. Hence the equations of motion ought to
follow from varying an action restricted to the external
universe. However, the boundary term in the derivation of the
Euler-Lagrange equations does not in general vanish on the stretched
horizon as it does at the boundary of spacetime. In order to obtain the correct
equations of motion, we must add to the external action a surface term
that cancels this residual boundary term. The membrane picture emerges
in interpreting the added surface term as electromagnetic and 
gravitational sources residing on the stretched horizon.

\section{Horizon Preliminaries}
In this section, we fix our conventions, first in words, then in equations. 
Through every point on the true horizon there exists a unique null
generator $l^a$ which we may parameterize by some regular time coordinate
whose normalization we fix to equal that of time-at-infinity.
Next, we choose a time-like surface just outside the true horizon. This is
the stretched horizon, $\cal{H}$, whose location we parameterize by 
$\alpha \ll 1$ so that
$\alpha \to 0$ is the limit in which the stretched horizon coincides
with the true horizon. We will always take this limit at the end of any
computation. Since many of the useful intermediate quantities will 
diverge as inverse powers of $\alpha$, we renormalize them by the 
appropriate power of $\alpha$. In
that sense, $\alpha$ plays the role of a regulator.

For our purposes, the principal reason
for preferring the stretched horizon over the true horizon is that the
metric on a time-like -- rather than null -- surface is nondegenerate, 
permitting one to write down a conventional action. Generically (in the
absence of horizon caustics), a one-to-one correspondence
between points on the true and stretched horizons is always possible
via, for example, ingoing 
null rays that pierce both surfaces (see \cite{pricethorne} for details).

We can take the stretched horizon to be the world-tube of a family of
time-like observers who hover just outside 
the true horizon. These nearly light-like 
``fiducial'' observers are pathological in that
they suffer an enormous proper acceleration and measure quantities that
diverge as $\alpha \to 0$. However, although we 
take the mathematical limit in
which the true and stretched horizons conflate, for physical purposes 
the proper distance of 
the stretched horizon from the true horizon need only be smaller 
than the length scale involved in a given measurement.
In that respect, the stretched horizon, although a
surrogate for the true horizon, is actually more fundamental
than the true horizon, since measurements at the stretched horizon constitute
real measurements that an external observer could make and report, whereas 
accessing any quantity measured at the true horizon would entail the observer's
inability to report back his or her results.

We take our fiducial observers to have world lines $U^a$, parameterized by
their proper time, $\tau$. The stretched horizon also possesses a space-like 
unit normal $n^a$ which for
consistency we shall always take to be outward-pointing. Moreover, we choose
the normal vector congruence on the stretched horizon to emanate outwards
along geodesics. We define $\alpha$ by requiring that $\alpha U^a 
\to l^a$ and 
$\alpha n^a \to l^a$; hence $\alpha U^a$ and $\alpha n^a$ are equal in
the true horizon limit. This is nothing more than the statement that the
null generator $l^a$ is both normal and tangential to the true horizon, which
is the defining property of null surfaces.
Ultimately, though, it will be this property that will be 
responsible for the dissipative behavior of the horizons.
The 3-metric, $h_{ab}$, on $\cal{H}$ can 
be written as a 4-dimensional tensor
in terms of the spacetime metric and the normal vector, so that
$h^a_b$ projects from the spacetime tangent space to the 3-tangent space. 
Similarly, we can
define the 2-metric, $\gm _{AB}$, of the space-like section of $\cal{H}$ 
to which
$U^a$ is normal, in terms of the stretched horizon 3-metric and $U^a$, thus
making a 2+1+1 split of spacetime. We denote the 4-covariant derivative
by $\del_a$, the 3-covariant derivative by $|a$, and the 2-covariant
derivative by $\|A$. For a vector in the stretched horizon, the
covariant derivatives are related by $h^c_d \del _c w^a = w^a_{|d} -
K^c_d w_c n^a$ where $K^a_b \equiv h^c_b \del_c n^a$ is the stretched
horizon's extrinsic curvature, or second fundamental form.
In summary,
\begin{equation}
l^2 = 0
\end{equation}
\begin{equation}
U^a = \lf \f{d}{d \tau} \rt ^a \; , \; \; U^2 = -1 \; , \; \; 
\lim_{\alpha\to\infty} \alpha U^a = l^a
\end{equation}
\begin{equation}
n^2 = + 1 \; , \; \; a^c = n^a \del _a n^c = 0 \; , \; \; 
\lim_{\alpha\to\infty} \alpha n^a = l^a
\end{equation}
\begin{equation}
h^a_b = g^a_b - n^a n_b \; , \; \; \gm^a_b = h^a_b + U^a U_b 
= g^a_b - n^a n_b + U^a U_b
\end{equation}
\begin{equation}
K^a_b \equiv h^c_b \del_c n^a \; , \; \; K_{ab} = K_{ba} \; , \; \; 
K_{ab} n^b = 0
\end{equation}
\begin{equation}
w^c \epsilon {\cal{H}} \Rightarrow h^c_d \del _c w^a = w^a_{|d} -
K^c_d w_c n^a \Rightarrow \del_c w^c = w^c _{|c} + w^c a_c = w^c _{|c}
\; . \label{div}
\end{equation}
The last expression relates the covariant divergence associated with $g_{ab}$
to the covariant divergence associated with $h_{ab}$.

For example, the Reissner-Nordstr\"{o}m solution has 
\begin{equation}
ds^2 = - \lf 1 - \f{2M}{r} + \f{Q^2}{r^2} \rt dt^2 + 
{\lf 1 - \f{2M}{r} + \f{Q^2}{r^2} \rt}^{-1} dr^2 + r^2 d \Omega ^2 \; ,
\end{equation}
so that a stretched horizon at constant $r$ would have
\begin{equation}
\alpha = {\lf 1 - \f{2M}{r} + \f{Q^2}{r^2} \rt}^{1 /2} \; ,
\end{equation}
\begin{equation}
U_a = - \alpha \lf d t \rt _a \; ,
\end{equation}
and
\begin{equation}
n_a = + \alpha ^{-1} \lf d r \rt _a \; .
\end{equation}

\section{Action Formulation}
To find the complete equations of motion by extremizing an action,
it is not sufficient to set the bulk
variation of the action to zero: one also needs to use the boundary 
conditions. Here we take our Dirichlet boundary conditions 
to be $\d \phi = 0$ at the boundary of spacetime,
where $\phi$ stands for any field. 

Now since the fields inside a black hole cannot have any 
classical relevance for an external observer,
the physics must follow from varying the part of the action restricted to the
spacetime outside the black hole. However, this external action is not
stationary on its own, because boundary conditions are fixed only at the
singularity and at infinity, but not at the stretched horizon. Consequently, we
rewrite the total action as

\begin{equation}
S_{\rm world} = \lf S_{\rm out} + S_{\rm surf} \rt + \lf S_{\rm in} - 
S_{\rm surf} \rt \; , \label{split}
\end{equation}
where now $\d S_{\rm out} + \d S_{\rm surf} \equiv 0$, 
which implies also that $\d S_{\rm in} - \d S_{\rm surf} = 0$. 
The total action has been broken down into two parts,
both of which are stationary on their own, and which do not require any
new boundary conditions.

The surface term, $S_{\rm surf}$, corresponds to sources, such as surface 
electric charges and currents for the Maxwell action, or surface stress
tensors for the Einstein-Hilbert action. The sources are fictitious: an
observer who falls through the stretched horizon will not
find any surface sources and, in fact, will not find any stretched horizon.
Furthermore, the field configurations inside the black hole will be 
measured by
this observer to be entirely different from those posited by the membrane
paradigm. On the other hand, for an external fiducial observer the
source terms are
a very useful artifice; their presence is consistent with all external
fields.
This situation is directly analogous to the method
of image charges in electrostatics, in which a fictitious charge
distribution is added to the system to implement, say, conducting boundary
conditions.
By virtue of the uniqueness of solutions to Poisson's equation with
conducting boundary conditions, the electric
potential on one -- and only one -- side of the boundary is guaranteed to
be the correct
potential. An observer who remains on that side of the boundary has no way
of telling through the fields alone whether they arise through the fictitious
image charges or through actual surface charges. The illusion is exposed only
to the observer who crosses the boundary to find that not only are there
no charges, but the potential on the other side of the boundary is
quite different from what it would have been had the image charges been real.

In the rest of this section, we shall implement Eq. (\ref{split}) 
concretely in important special cases.

\subsection{The Electromagnetic Membrane}
The external Maxwell action is
\begin{equation}
S_{\rm out}[A_a] = \int d^4 x \rg \lf - \f {1}{16 \pi} F^2 + J \cdot A \rt \; ,
\end{equation}
where $F$ is the electromagnetic field strength.
Under variation, we obtain the inhomogeneous Maxwell equations
\begin{equation}
\del_b F^{ab} = 4 \pi J^a \; ,
\end{equation}
as well as the boundary term
\begin{equation}
\f {1} {4 \pi} \int d^3 x \rh F^{ab} n_a \d A_b \; ,
\end{equation}
where $h$ is the determinant of the induced metric, 
and $n^a$ is the outward-pointing space-like unit normal to the
stretched horizon. We need to cancel this term. Adding the surface term
\begin{equation}
S_{\rm surf}[A_a] = + \int d^3 x \rh j_s \cdot A \; ,	\label{charge}
\end{equation}
we see that we must have
\begin{equation}
j_s^a = + \f {1} {4 \pi} F^{a b} n_b \; .	\label{jmaxwell}
\end{equation}
The surface 4-current, $j_s^a$,
has a simple physical interpretation. We see 
that its time-component is a surface charge, $\sigma$, 
that terminates the normal component of the electric 
field just outside the membrane, while the spatial
components, $\vec{j_s}$, form a surface current 
that terminates the tangential component
of the external magnetic field:
\begin{equation}
E_{\perp} = -U_a F^{ab} n_b = 4 \pi \sigma
\end{equation}
\begin{equation}
\vec{B}_{\|}^A = \epsilon^A_B \gm^B_a F^{ab} n_b = 4 \pi \lf \vec{j}_s 
\times \hat{n} \rt ^A	\; . \label{B}
\end{equation}
It is characteristic of the membrane paradigm that $\sigma$ and $\vec{j}_s$ are
{\em local} densities, so that the total charge on the black hole is the
surface integral
of $\sigma$ over the membrane, taken at some constant universal time.
This is in contrast to the total
charge of a Reissner-Nordstr\"{o}m black hole, which is a 
global characteristic that can be defined by an integral at spatial infinity.

{}From Maxwell's equations and Eq. (\ref{jmaxwell}), we obtain a continuity 
equation for
the membrane 4-current which, for a stationary hole, takes the form
\begin{equation}
\f {\pl \sigma}{\pl \tau} + \vec{\del}_2 \cdot \vec{j}_s  = - J^n \; ,
\end{equation}
where $\vec{\del}_2 \cdot \vec{j}_s \equiv {\lf \gm^A_a j_s^a \rt}_{\|A}$ 
is the two-dimensional divergence of 
the membrane surface current, and $- J^n = -J^a n_a$ is the amount of charge
that falls into the hole per unit area per unit proper time, $\tau$.
Physically, this equation expresses local charge conservation in that 
any charge that falls into the black hole can be regarded as remaining on 
the membrane: the membrane is impermeable to charge.

The equations we have so far are sufficient to determine the fields outside
the horizon, given initial conditions outside the horizon.
A plausible requirement for initial conditions {\em at} the horizon is that the
fields measured by freely falling observers (FFO's) at the stretched horizon 
be finite. 
There being no curvature singularity at the horizon, inertial observers
who fall through the horizon should detect nothing out of the ordinary. In
contrast, the fiducial observers (FIDO's) who make measurements at the membrane
are infinitely accelerated. Their measurements, subject to infinite
Lorentz boosts, are singular. For the electromagnetic fields we
have, with $\gm$ the Lorentz boost and using orthonormal coordinates,
\begin{eqnarray}
E_{\theta}^{\rm FIDO} \approx  \gm \lf E_{\theta}^{\rm FFO} - B_ 
{\phi}^{\rm FFO} \rt \; , \; \; &
B_{\phi}^{\rm FIDO} \approx  \gm \lf B_{\phi}^{\rm FFO} - E_
{\theta}^{\rm FFO} \rt \; , \\
B_{\theta}^{\rm FIDO} \approx  \gm \lf B_{\theta}^{\rm FFO} - E_
{\phi}^{\rm FFO} \rt \; , \; \; &
E_{\phi}^{\rm FIDO} \approx  \gm \lf E_{\phi}^{\rm FFO} - B_
{\theta}^{\rm FFO} \rt \; ,
\end{eqnarray}
or, more compactly,
\begin{equation}
\vec{E}^{\rm FIDO}_{\|} = \hat{n} \times \vec{B}^{\rm FIDO}_{\|} \; .	
\label{reg}
\end{equation}
That is, the regularity condition states that all radiation in the 
normal direction is ingoing; a black hole acts as a perfect absorber.
Combining the regularity condition with Eq. (\ref{B}) and dropping the FIDO
label, we arrive at
\begin{equation}
\vec{E}_{\|} = 4 \pi \vec{j}_s	\; .
\end{equation}
That is, black holes obey Ohm's law with a surface resistivity
of $\rho = 4 \pi \approx 377 \; \Omega$. Furthermore, the Poynting flux is
\begin{equation}
\vec{S} = \f {1} {4 \pi} \lf \vec{E} \times \vec{B} \rt = - j_s^2 \rho \, 
\hat{n} \; .
\end{equation}
We can integrate this over the black hole horizon at some fixed
time. However, for a generic stretched horizon, we cannot
time-slice using fiducial time as different fiducial 
observers have clocks that do not necessarily remain synchronized. Consequently
we must use some other time for slicing purposes, such as the time at
infinity, and then include in the integrand a (potentially
position-dependent) factor to convert the locally measured energy flux
to one at infinity. With a clever choice of the stretched horizon,
however, it is possible to arrange that all fiducial observers have
synchronized clocks. In this case, two powers of $\alpha$, which is
now the lapse, are included in the integrand. 
Then, for some given universal time, $t$, the power radiated into
the black hole, which is also the rate of increase of the black hole's
irreducible mass, is given by
\begin{equation}
\f {d M_{\rm irr}} {d t} = - \int \alpha^2 \vec{S} \cdot d \vec{A} = 
+ \int \alpha^2 j_s^2 \rho \, dA \; .
\end{equation}
That is, black holes obey the Joule heating law, the same law that also
describes the dissipation of an ordinary Ohmic resistor.

\subsection{The Gravitational Membrane}
We turn now to gravity. The external Einstein-Hilbert action is
\begin{equation}
S_{\rm out}[g^{ab}] = \f {1} {16 \pi} \int d^4 x \rg R + \f {1} {8 \pi} \oint
d^3 x \sqrt {\pm h} \, K + S_{\rm matter} \; ,
\end{equation}
where $R$ is the Ricci scalar, and $K$ is the trace of the extrinsic 
curvature, and where for convenience we have chosen the
field variable to be the inverse metric $g^{ab}$.
The surface integral of $K$ is only over the outer boundary of spacetime, 
and not over the stretched horizon. It is
required in order to obtain the Einstein equations because the Ricci scalar
contains second order derivatives of $g_{ab}$. When this action is varied,
the bulk terms give the Einstein equations
\begin{equation}
R_{ab} - \f {1}{2} g_{ab} R = 8 \pi T_{ab} \; . \label{ein}
\end{equation}
We are interested, however, in the interior boundary term. This comes from the
variation of the Ricci tensor. We note that
\begin{equation}
g^{ab} \d R_{ab} = \del ^a \left [ \del ^b \lf \d g_{ab} \rt - 
g^{cd} \del _a \lf \d g_{cd} \rt \right ] \; ,
\end{equation}
where $\d g_{ab} = - g_{ac} g_{bd} \d g^{cd}$. Gauss' theorem now gives
\begin{equation}
\int d^4 x \rg \lf g^{ab} \d R_{ab} \rt = - \int d^3 x \rh n^a h^{bc} \left [ 
\del_c \lf \d g_{ab} \rt - \del_a \lf \d g_{bc} \rt \right ] \; ,
\end{equation}
where the minus sign arises from choosing $n^a$ to be 
outward-pointing. Applying the Leibniz rule, we can rewrite this as
\begin{eqnarray}
\int d^4 x \rg \lf g^{ab} \d R_{ab} \rt & = & \int d^3 x \rh h^{bc} \left [ 
\del _a \lf n^a \d g_{bc} \rt 
{} - \d g_{bc} \del _a \lf n^a \rt \right. \nonumber \\
& & \indent \indent \; \; \left. 
{} - \del _c \lf n^a \d g_{ab} \rt
+ \d g_{ab} \del _c \lf n^a \rt \right ] \; .	\label{long}
\end{eqnarray}

Now, in the limit that the stretched horizon approaches the null horizon, the
first and third terms on the right-hand side vanish:
\begin{equation}
\int d^3 x  \rh h^{bc} \left [ \del _a \lf n^a \d g_{bc} \rt 
{} - \del _c \lf n^a \d g_{ab} \rt \right ] = 0 \; .	\label{zero}
\end{equation}
A proof of this identity is given in the Appendix. With $K^{ba} =
h^{bc} \del_c n^a$, the variation of the external action is 
\begin{equation}
\d S_{\rm out}[g^{ab}] = {1 \over 16 \pi} \int d^3 x \rh 
\lf K h_{ab} - K_{ab} \rt \d g^{ab} \; .	\label{var}
\end{equation}
Since the expression in parentheses contains only stretched horizon tensors,
the normal vectors in the variation $\d g^{ab} = \d h^{ab} + 
\d n^a n^b + n^a \d n^b$ contribute nothing.
As in the electromagnetic case, we add a surface source term to the action
to cancel this residual boundary term. The variation of the required term 
can therefore be written as
\begin{equation}
\d S_{\rm surf}[h^{ab}] = - \f {1} {2} \int d^3 x \rh t_{s \, ab} \d
h^{ab} \; . \label{cancel}
\end{equation}
We shall see later that this variation is integrable; i.e., an
action with this variation exists. Comparison with Eq. (\ref{var})
yields the membrane stress tensor
\begin{equation}
t_s^{ab} = + \f {1} {8 \pi} \lf K h^{ab} - K^{ab} \rt \; . \label{surf}
\end{equation}
Now just as a surface charge produces a discontinuity in the normal component
of the electric field, a surface stress term creates a discontinuity in the
extrinsic curvature. The relation between the discontinuity and the source
term is given by the Israel junction condition \cite{junction,mtw},
\begin{equation}
t_s^{ab} = \f {1} {8 \pi} \lf [K] h^{ab} - [K]^{ab} \rt	\; ,
\end{equation}
where $[K] = K_+ - K_-$ is the difference in the extrinsic curvature of the
stretched horizon between its embedding in the external universe and its
embedding in the spacetime internal to the black hole.
Comparing this with our result for the membrane stress tensor, Eq. 
(\ref{surf}), we see that
\begin{equation}
K_- ^{ab} = 0 \; ,
\end{equation}
so that the interior of the stretched horizon molds itself into flat space.
The Einstein equations, Eq. (\ref{ein}), can be rewritten
via the contracted Gauss-Codazzi equations \cite{mtw} as
\begin{equation}
t^{ab}_{s \; \: |b} = - h^a_c T^{cd} n_d \; .	\label{gcod}
\end{equation}

Equations (\ref{surf}) and (\ref{gcod})
taken together imply that the stretched horizon can be thought of as a fluid
membrane, obeying the Navier-Stokes equation. To see this, recall that
as we send $\alpha$ to zero, both $\alpha U^a$ and $\alpha n^a$ approach $l^a$,
the null generator at the corresponding point on the true horizon. 
Hence, in this limit we can equate $\alpha U^a$ and $\alpha n^a$, permitting us
to write the relevant components of $K^a_b$, in terms of the surface
gravity, $g$, and the extrinsic curvature, $k^A_B$,
of a space-like 2-section of the stretched horizon:
\begin{equation}
U^c \del _c n^a \to \alpha ^{-2} l^c \del _c l^a 
\equiv \alpha ^{-2} g_H l^a
\Rightarrow K_U^U = - g \; , \; \;
K^A_U = \gm^A_a K^a_b U^b = 0 \; ,
\end{equation}
where $g_H \equiv \alpha g$ is the renormalized surface gravity at 
the horizon, and
\begin{equation}
\gm_A ^c \del _c n^b \to \alpha ^{-1} \gm_A^c \del _c l^b
\Rightarrow K_A^B = \gm_A^a K_a^b \gm_b^B = \alpha ^{-1} k_A^B \; ,
\end{equation}
where $k_{AB}$ is the extrinsic curvature of a space-like 2-section of 
the true horizon,
\begin{equation}
k_{AB} \equiv \gm_A^d l_{B \| d} = {1 \over 2} \pounds _{l^a} \gm _{AB} \; ,
\end{equation}
where $\pounds _{l^a}$ is the Lie derivative in the direction of $l^a$. We 
can decompose $k_{AB}$ into a traceless part and a trace, $k_{AB} = 
\sigma_{AB} + \f {1} {2} \gm_{AB} \theta$, where $\sigma_{AB}$ is the shear
and $\theta$ the expansion of the world lines of nearby horizon surface
elements. Then
\begin{equation}
t_s^{AB} = \f {1} {8 \pi} \left [ - \sigma ^{AB} + \gm^{AB} \lf \f {1} {2}
\theta + g \rt \right ] \; .	\label{stress}
\end{equation}
But this is just the equation for the stress of a two-dimensional viscous
Newtonian fluid \cite{LLFM} with pressure $p = g / 8 \pi$, shear viscosity 
$\eta = 1 / 16 \pi$, and bulk viscosity $\zeta = - 1 / 16 \pi$.
Hence we may identify the horizon with a two-dimensional dynamical fluid, or
membrane. Note that, unlike ordinary fluids, 
the membrane has negative bulk viscosity. This would ordinarily
indicate an instability against generic perturbations triggering 
expansion or contraction. It can be regarded as reflecting
a null hypersurface's natural tendency to expand or contract
\cite{damour82}. Below we shall show how for the horizon this particular 
instability is replaced with a different kind of instability.

Inserting the $A$-momentum density ${t_s}^b_a \gm ^a_A U_b = 
t^{\; U}_{s \; \; A} \equiv \pi _A$ 
into the Einstein equations, Eq. (\ref{gcod}), we arrive at 
the Navier-Stokes equation
\begin{equation}
\pounds _{\tau} \pi _A = - \del _A p + \zeta \del _A \theta + {\rm 2} \eta
\sigma ^B _{A \; \; \| B} - T^n_A \; ,
\end{equation}
where $\pounds _{\tau} \pi _A =  \pl \pi _A / \pl \tau$ is the 
Lie derivative (which is the general-relativistic equivalent of the convective 
derivative) with respect to proper time, and
$- T^n_A = - \gm ^a_A T^c_a n_c$ is the flux of 
$A$-momentum into the black hole.

Inserting the $U$-momentum (energy) density $t^{\; a}_{s \; \; b} U_a U^b 
\equiv \Sigma = - \theta / 8 \pi$ gives
\begin{equation}
{\pounds} _{\tau} \Sigma + \theta \Sigma = - p \theta + \zeta \theta ^{\rm 2}
+ {\rm 2} \eta \sigma _{AB} \sigma ^{AB} + T^a_b n_a U^b \; , \label{energy}
\end{equation}
which is the focusing equation for a null geodesic congruence \cite{focus}. 
We might now suspect that if the analogy with fluids extends to 
thermodynamics, then Eq. (\ref{energy}), as the equation of energy 
conservation, must be the heat transfer equation \cite{LLFM} for 
a two-dimensional 
fluid. Writing the expansion of the fluid in terms of the area, $\Delta A$, of
a patch,
\begin{equation}
\theta = \f {1} {\Delta A} \f {d \Delta A} {d \tau} \; ,
\end{equation}
we see that we can indeed rewrite Eq. (\ref{energy}) as the heat 
transfer equation (albeit with an additional relativistic term on the left)
\begin{equation}
T \lf \f {d \Delta S} {d \tau} - \f {1} {g} \f {d^2 \Delta S} {d \tau ^2} \rt =
\lf \zeta \theta ^2 + 2 \eta \sigma _{AB} \sigma ^{AB} + T^a_b n_a U^b \rt 
\Delta A \; ,	\label{heat}
\end{equation}
with $T$ the temperature and $S$ the entropy, provided that the 
entropy is given by
\begin{equation}
S = \eta \f {k_B} {\hbar} A \; , 	\label{ent}
\end{equation}
and the temperature by
\begin{equation}
T = \f {\hbar} {8 \pi k_B \eta} g \; ,	\label{temp}
\end{equation}
where $\eta$ is some proportionality constant. 

Thus, the identification of the horizon
with a fluid membrane can be extended to the thermodynamic domain.
Nonetheless, the membrane is an unusual fluid.
The focusing equation itself, Eq. (\ref{energy}), is 
identical in form to the
equation of energy conservation for a fluid. However, because the
energy density, $\Sigma$, is proportional to the expansion, $\Sigma
= - \theta / 8 \pi$, one obtains a nonlinear first-order
differential equation for $\theta$ which has no counterpart for ordinary
fluids. The crucial point is that, owing to the black hole's
gravitational self-attraction, the energy density is negative, and the
solution to the differential equation represents a horizon that 
grows with time.
For example, the source-free solution with a time-slicing for which the
horizon has constant surface gravity is
\begin{equation}
\theta \lf t \rt = \f {2 g} {1 + \lf \f{2 g}{\theta \lf t_0 \rt} - 1 \rt 
e^{g \lf t_0 - t \rt }} \; .
\end{equation}
Because of the sign of the exponent, this would represent an ever-expanding
horizon if
$\theta \lf t_0 \rt$ were an initial condition; the area of the
horizon, which is related to $\theta$ by $\theta = (d / d \tau ) \ln
\sqrt{\gamma} \,$, expands exponentially with time. To avoid this runaway, 
one must impose ``teleological boundary conditions'' (that is, final 
conditions) rather than initial conditions. 
Hence, the horizon's growth is actually
acausal; the membrane expands to intercept infalling matter that is yet
to fall in \cite{mempar}. 
This is because the membrane inherits the global character 
of the true horizon: the stretched horizon covers
the true horizon whose location can only be determined by tracking null
rays into the infinite future. 
In fact, the left-hand side of the heat transfer equation,
Eq. (\ref{heat}), is of the same form as that of an electron subject
to radiation reaction; the acausality of the horizon is therefore
analogous to the pre-acceleration of the electron.

At this
classical level, using only the equations of motion, the parameter 
$\eta$ in Eq. (\ref{ent})
is undetermined. However, because we have an action we hope to do
better, since the normalization in the path integral is now fixed.
By means of a Euclidean path integral, we should actually be able to
derive the Bekenstein-Hawking entropy, including the coefficient $\eta$, 
from the membrane action. We do this in the next chapter.

\subsection{The Axidilaton Membrane}
Another advantage of the action formulation is that it is easily 
generalized to arbitrary fields. For example, we can extend
the membrane paradigm to include the basic fields of quantum gravity.
Here we use the tree-level effective action obtained from
string theory after compactification to four macroscopic dimensions.
This action is a generalization of the
classical Einstein-Hilbert-Maxwell action to which it reduces when
the axidilaton, $\lm$, is set to $i / 16 \pi$. The action is
\begin{equation}
S[ \lm , \cc \lm, A_a, g_{ab} ] = \int d^4 x \rg \lf \f{R}{16 \pi} 
{}-  \f { | \pl \lm |^2}{2 \l2 ^2} 
+  \f {i}{4} \lf\lm F_+ ^2 - \cc \lm F_- ^2 \rt \rt \; ,
\end{equation}
where $R$ is the four-dimensional Ricci curvature scalar,
$F_{\pm} \equiv F \pm i \tilde F$ are the self- and anti-self-dual 
electromagnetic field strengths, and
$\lm \equiv \lm _1 + i \l2 = a + i e^{-2 \phi}$ is the axidilaton, with
$a$ the axion and $\phi$ the dilaton. Solutions to
the equations of motion arising from this action include electrically
(Reissner-Nordstr\"{o}m) and magnetically charged black holes 
\cite{gibbonsmaeda,ghs}, as well as their duality-rotated cousins, dyonic 
black holes \cite{shaperefw},which carry both electric and magnetic charge.

The equations of motion are

\begin{equation}
\del _a \lf \f {\pl ^a \lm}{\l2 ^2} \rt + i \f {| \pl \lm |^2}{\l2 ^3} 
{} - \f {i}{2} F_- ^2 = 0
\end{equation}
and
\begin{equation}
\del _a \lf \lm F_+^{ab} - \cc \lm F_-^{ab} \rt = 0 \; ,
\end{equation}
besides the Einstein equations.

As before, we require the external action to vanish on its own. Integration
by parts on the axidilaton kinetic term leads to a variation at the boundary,
\begin{equation}
\int d^3 x \rh \left [
 \d \lm \lf \f {n_a \pl ^a \cc \lm} {2 \l2 ^2} \rt + \d \cc 
\lm \lf \f {n_a \pl ^a \lm} {2 \l2 ^2} \rt \right ] \; ,
\end{equation}
where $n^a$ is again chosen to be outward-pointing.
To cancel this, we add the surface term
\begin{equation}
S_{\rm surf} = \int d^3 x \rh \lf \lm \cc q + \cc \lm q \rt \; ,
\end{equation}
so that
\begin{equation}
q = - \f {n_a \pl ^a \lm} {\l2 ^2} \; . \label{q}
\end {equation}

To interpret this, we note that the kinetic term in $\lm$ is invariant under 
global $SL \lf 2, I\!\!R \rt$ transformations of the form
\begin{equation}
\lm \to \f {a \lm + b}{c \lm + d} \; , \; \; ad - bc = 1 \; ,
\end{equation}
which are generated by Peccei-Quinn shifts, $\lm_1 \to \lm_1 + b$,
and duality transformations, $\lm \to - 1 / \lm$. The Peccei-Quinn
shift of the axion can be promoted to a classical local symmetry to
yield a N\"other current:
\begin{equation}
J^a_{P-Q} = - \f{1}{2 \lm_2^2} \lf \pl ^a \lm + \pl ^a \cc \lm \rt \; .
\end{equation}
Therefore, under a Peccei-Quinn shift,
\begin{equation}
\d S_{\rm surf} = \int d^3 x \rh \d \lm \lf q + \cc q \rt
= \int d^3 x \rh \d \lm \lf n_a J^a_{P-Q} \rt \; .
\end{equation}
The sum of the $q$ and $\cc q$ terms induced at the 
membrane, Eq. (\ref{q}), is the normal component of the Peccei-Quinn current.
Hence, at the membrane,
\begin{equation}
\lf h^a_b J^b_{P-Q} \rt _{|a} = - F \tilde F - \del_a 
\left [ \lf q + \cc q \rt n^a \right ] \; .
\end{equation}
That is, the membrane term $\del_a \left [ \lf q + \cc q \rt n^a \right ]$ 
augments the dyonic $F \tilde F$ term as a source for 
the three-dimensional Peccei-Quinn
current, $h^a_b J^b_{P-Q}$, at the membrane.

The membrane is again dissipative with the Peccei-Quinn charge accounting
for the dissipation in the usual $\alpha \to 0$ limit. The local rate of
dissipation is given by the bulk stress tensor at the membrane:
\begin{equation}
T_{ab} U^a n^b = \f{1}{16 \pi} \f {\pl_a \lm \pl_b \cc \lm + \pl_a \cc \lm 
\pl_b \lm} {2 \lm_2 ^2} U^a n^b \to \f{\lm_2^2 |q|^2}{16 \pi} \; .
\end{equation}

In addition, the presence of the axidilaton affects the 
electromagnetic membrane.
(The gravitational membrane is unaffected since the surface terms come from
the Ricci scalar which has no axidilaton factor.) The electromagnetic
current is now

\begin{equation}
j_s^a = - 2 i \lf \lm F_+ ^{ab} - \cc \lm F_- ^{ab} \rt n_b \; .
\end{equation}
The surface charge is therefore
\begin{equation}
\sigma = 4 \lf \lm _2 E_\perp + \lm _1 B_\perp \rt \; ,
\end{equation}
and the surface current is
\begin{equation}
\vec{j}_s = 4 \lf \lm _2 \hat{n} \times \vec{B}_{\|} - \lm _1 \hat{n} 
\times \vec{E}_{\|} \rt \; ,
\end{equation}
which, by the regularity of the electromagnetic field, Eq. (\ref{reg}),
satisfies
\begin{equation} 
\lf \begin{array}{c} j_s^{\theta} \\ j_s^{\phi} \end{array} \rt =
4 \lf \begin{array}{cc} \lm _2 & \lm _1 \\ - \lm _1 & \lm _2 \end{array} \rt
\lf \begin{array}{c} E^{\theta} \\ E^{\phi} \end{array} \rt \; .
\end{equation}
The conductivity is now a tensor. When the axion is absent, the resistivity
is
\begin{equation}
\rho = \f {1} {4 \lm _2} \; .
\end{equation}
The inverse dependence on $\lm_2$ is to be expected on dimensional
grounds. The pure dilaton action can be
derived from Kaluza-Klein compactification of pure gravity in five dimensions,
where the fifth dimension is curled into a circle of radius $e^{-2 \phi} = 
\lm _2$. In five dimensions, with $c \equiv 1$, resistance (and hence
resistivity for a two-dimensional resistor such as the membrane) has 
dimensions of inverse length.
Using the regularity condition, Eq. (\ref{reg}), the rate of
dissipation, for a stretched horizon defined to have uniform lapse
$\alpha$ with respect to time at infinity, $t$, is
\begin{equation}
\f{d M_{\rm irr}}{d t} = - \int \alpha^2 \vec{S} \cdot d \vec{A} = 
\int 4 \alpha^2 \lm_2 E^2_{\|} dA  
= \int \alpha^2 \f{\lm_2}{4 |\lm|^4} \vec{j_s}^2 dA \; ,
\end{equation}
which is the Joule heating law in the presence of an axidilaton.

\section{Dissipation}

Given that the bulk equations of motion are manifestly symmetric under 
time-reversal,
the appearance of dissipation, as in Joule heating and fluid viscosity, might 
seem mysterious, all the more so since it has been derived from an action.

The procedure, described here, of restricting the action to some region and 
adding surface  terms on the boundary of the region cannot be applied with
impunity to any arbitrary region: a black hole is special. 
This is because the region
outside the black hole contains its own causal past; an observer who remains
outside the black hole is justified in neglecting (indeed, is unaware of)
events inside. However, even ``past sufficiency'' does not adequately
capture the requirements for our membrane approach. For 
instance, the past light cone of a spacetime point obviously
contains its own past, but an observer in this light cone must eventually
leave it. Rather, we define the notion of a future dynamically closed set:
\begin{quote}
A set $S$ in a time-orientable globally hyperbolic 
spacetime $(M, g_{ab})$ is {\em future dynamically closed} if
$J^-(S) = S$, and if, for some foliation
of Cauchy surfaces $\Sigma _t$ parameterized by the values
of some global time function, we have that $\forall \; t_0 \; \forall \; 
p \, \epsilon \, \lf S \cap \Sigma _{t_0} \rt \; \forall \; (t > t_0) \;
\exists \; q \, \epsilon \,  \lf I^+(p) \cap S \cap \Sigma _t \rt$.
\end{quote}
That is, $S$ is future dynamically closed if it contains
its own causal past and if from every point in $S$ it is possible for
an observer to remain in $S$.
Classically, the region outside the true horizon of a black hole is
dynamically closed. So too is the region on one side of a null plane in flat
space; this is just the infinite-mass limit of a black hole. 
The region outside the stretched horizon is strictly speaking {\em not} 
dynamically closed since a signal originating in the thin
region between the stretched horizon and the true horizon
can propagate out beyond the stretched horizon. However, in the limit
that the stretched horizon goes to the true horizon, $\alpha \to 0$,
this region becomes vanishingly thin so that in this limit, which is in any
case assumed throughout, we are justified in restricting the action.

The breaking of time-reversal symmetry comes from the definition of the
stretched horizon; the region exterior to the black hole does not remain
future dynamically closed under time-reversal. In other words, we have divided
spacetime into two regions whose dynamics are derived from two different
simultaneously vanishing actions, $\d \lf S_{\rm out} + S_{\rm surf} \rt
= \d \lf S_{\rm in} - S_{\rm surf} \rt = 0$. 
Given data on some suitable achronal subset we can, for the exterior 
region, predict the future but not the entire past, while, inside the
black hole, we can ``postdict'' the past but cannot determine the
entire future. Thus, our choice of the horizon as a boundary implicitly
contains the irreducible logical requirement for dissipation, that is,
asymmetry between past and future.

Besides the global properties that logically permit one to write down a
time-reversal asymmetric action, there is also a local property of
the horizon which is the proximate cause for dissipation, namely that the 
normal
to the horizon is also tangential to the horizon. Without this crucial
property -- which manifests itself as the regularity condition, or
the identification of the stretched horizon
extrinsic curvature with intrinsic properties of the true horizon -- there
would still be surface terms induced at the stretched horizon, but no
dissipation.

The regularity condition imposed at the boundary is not an operator
identity, but a statement about physical states: all tangential
electromagnetic fields as measured by a fiducial observer must be
ingoing. Such a statement is not rigorously true. For any given value
of $\alpha = \lf 1 - 2M / r \rt ^{1 / 2}$, there is a maximum wavelength,
$\lm_{\rm max}$, for outgoing modes that are invisible to the observer:
\begin{equation}
\lm_{\rm max} = \f {r - 2M}{\lf 1 - 2M / r \rt ^{1 / 2}} \rightarrow 2
M \alpha \; .
\end{equation}
Dissipation occurs in the membrane paradigm because the finite but
very high-frequency modes that are invisible to the fiducial observer 
are tacitly assumed not to exist. The regularity condition amounts to a
coarse-graining over these modes.

We conclude this section with an illustration of the intuitive advantage of 
the membrane paradigm. It is a famous result that the external state 
of a stationary black hole, quite unlike that of other macroscopic 
bodies, can be completely characterized by only four quantities: the 
mass, the angular momentum, and the electric and magnetic charges. 
That such a ``no-hair theorem'' \cite{nohair} should hold is certainly 
not immediately apparent from other black hole viewpoints.
In the membrane picture, however, we can see this fairly easily. For example,
an electric dipole that falls into the black hole can now be considered 
as merely two opposite charges incident on a conducting surface. 
The charges cause a current to flow and the current eventually 
dissipates; in the same way, all higher multipole moments are 
effaced. Similarly, the gravitational membrane obeys the Navier-Stokes 
equation, which is also dissipative; higher moments of an infalling 
mass distribution are thus obliterated in the same way. The only 
quantities that survive are those protected
by the conservation laws of energy, angular momentum, and electric
and magnetic charge. While far from an actual proof, this is at least a 
compelling and physically appealing argument for why black holes have 
only four ``hairs.''

\section{Hamiltonian Formulation}
The equations of motion can equally well be derived within a 
Hamiltonian formulation. This involves first singling out a global time
coordinate, $t$, for the external universe, which is then sliced 
into space-like surfaces, $\Sigma _t$, of constant 
$t$. We can write, in the usual way,
\begin{equation}
t^a \equiv \lf \f{d}{dt} \rt ^a = \alpha U^a - v^a \; ,
\end{equation}
where $U^a$ is the unit normal to $\Sigma _t$, $U^2 = -1$, and $\alpha$ and
$- v^a$ are Wheeler's lapse and shift, respectively, with $v^a = dx^a / dt$
the ordinary 3-velocity of a particle with world-line $U^a$.
For convenience we choose the stretched horizon to be a surface of constant
lapse so that $\alpha$, which goes to zero at the true horizon, serves
as the stretched horizon regulator. The external 
Hamiltonian for electrodynamics, 
obtained from the Lagrangian via a Legendre transform and written in
ordinary three-dimensional vector notation, is
\begin{equation}
H_{\rm out}[\phi, \vec{A}, \vec{\pi}] = \f{1}{4 \pi} \int_{\Sigma_t} 
d^3 x \sqrt{^3 g} \lf \f{1}{2} \alpha
\lf \vec{E} \cdot \vec{E} + \vec{B} \cdot \vec{B} \rt + \vec{v} \cdot
\lf \vec{E} \times \vec{B} \rt - \phi \lf \vec{\del} \cdot \vec{E} \rt
\rt \; , 	\label{ham}
\end{equation}
where $^3 g_{ab}$ is the 3-metric on $\Sigma_t$, 
$\phi \equiv -A_a t^a$ is the scalar potential, $\vec{A}_a \equiv 
{^3 g_{a}^{\; b} A_b}$
is the three-dimensional vector potential, and $\vec{\pi}^a \equiv - 
\sqrt{^3 g} \vec{E}^a$ its canonical momentum conjugate. Note that
$E^a = F^{ab} U_b$ is the co-moving electric field; $\vec{E}$ and $\vec{B}$
above refer to the fields measured by a fiducial observer with world-line
$U^a$.
Finally, the scalar potential is nondynamical; its
presence in the Hamiltonian serves to enforce Gauss' law as a constraint.
The equations of motion are now determined by Hamilton's equations and the
constraint:
\begin{equation}
\f{\d H}{\d \vec{\pi}} = \dot{\vec{A}} \; , \; \; \f{\d H}{\d \vec{A}} 
= - \dot{\vec{\pi}} \; , \; \; \f{\d H}{\d \phi} = 0 \; .
\end{equation}
In the bulk these equations are simply Maxwell's equations but, because 
of the inner boundary, the usually discarded surface terms that arise 
during integration by parts now need to be canceled.
It is easy to show then that the above equations hold only
if additional surface terms are added to the Hamiltonian:
\begin{equation}
H = H_{\rm out} - \int d^2 x \sqrt{\gm} \, j_s \cdot A \; .
\end{equation}
For Maxwell's equations to be satisfied in the bulk, the surface terms
are once again the surface charges and currents necessary to terminate
the normal electric and tangential magnetic fields at the stretched horizon.
Thus, the membrane paradigm is recovered.

However, it is perhaps more interesting to proceed in a slightly different
fashion. Instead of adding new terms, we can use the external Hamiltonian
to prove the validity of a principle of minimum heat production.
Such a principle, which holds under rather general circumstances for
stationary dissipative systems, holds for black holes also in
slightly nonstationary situations.

Now the time derivative of the external Hamiltonian is not zero, again
because of the inner boundary. We can use Hamilton's
equations to substitute expressions for the time derivative of the field
and its momentum conjugate. Hamilton's equations are
\begin{equation}
\dot{\vec{A}} = - \alpha \vec{E} + \vec{v} \times \vec{B} - \vec{\del} \phi
\label{adot}
\end{equation}
\begin{equation}
\dot{\vec{E}} = \vec{\del} 
\times \lf \alpha \vec{B} + \vec{v} \times \vec{E} \rt \; ,
\end{equation}
so that, making repeated use of the vector identity
\begin{equation}
\vec{\del} \cdot \lf \vec{V} \times \vec{W} \rt = \vec{W} \cdot \lf \vec{\del}
\times \vec{V} \rt - \vec{V} \cdot \lf \vec{\del} \times \vec{W} 
\rt \; ,
\end{equation}
we find that the energy loss is
\begin{equation}
{}- \dot{H} = - \f{1}{4 \pi} \int d^2 x \sqrt{\gm} \, 
\left [ \hat{n} \cdot \lf \alpha \vec{E}_{\|} \times \alpha \vec{B}_{\|} \rt
+ \vec{v} \cdot \lf E_{\perp} \alpha \vec{E}_{\|} 
+ B_{\perp} \alpha \vec{B}_{\|} \rt \right ]	\; .	\label{dHdt}
\end{equation}
So far, we have used only Hamilton's equations. It
remains, however, to implement the constraint. Hence we may regard $- \dot{H}$
as a functional of the Lagrange multiplier, $\phi$. We therefore have
\begin{equation}
{}- \f{\d \dot{H}}{\d \phi} = - \f{d}{dt} \f{\d H}{\d \phi} = 0 \; .
\end{equation}
That is, the equations of motion follow from minimizing the rate of
mass increase of the black hole with respect to the scalar potential. This
is an exact statement; we now show that this reduces to a minimum heat
production principle in a quasi-stationary limit.
Now we note that the first law of black hole thermodynamics
allows us to decompose the mass change into irreducible and rotational parts:
\begin{equation}
\f{d M}{d t} = \f{d Q}{d t} + \Omega_H \f{d J}{d t} \; ,
\end{equation}
where $\Omega_H$ is the angular velocity at the horizon, and $J$ is the
hole's angular momentum. Since $|\vec{v}| \to \Omega_H$ at the horizon, we see
that the second term on the right in Eq. (\ref{dHdt}) corresponds to the
torquing of the black hole. When this is small, we may approximate the
mass increase as coming from the first, irreducible term. Hence, in the
quasi-stationary limit, for a slowly rotating black hole, 
the black hole's rate of 
mass increase is given by the dissipation of external energy.
Invoking the regularity condition, Eq. (\ref{reg}), then gives
\begin{equation}
D[\phi] = + \f{1}{4 \pi} \int d^2 x \sqrt{\gm} \, \lf \alpha \vec{E}_{\|} \rt 
^2 \; , \; \; \f{\d D}{\d \phi} = 0 \; ,
\end{equation}
where $\alpha \vec{E}_{\|}$ is given by Eq. (\ref{adot}).
This is the principle of minimum heat production: minimizing the dissipation
functional leads to the membrane equation of motion.

We observe that we could have anticipated this answer. 
The numerical value of the Hamiltonian 
is the total energy of the system as measured at spatial infinity (assuming 
an asymptotically flat spacetime). The time derivative is then
simply the rate, as measured by the universal time of distant observers,
that energy changes. The rate of decrease of energy is the integral of the
Poynting flux as measured by local observers, multiplied by two powers of
$\alpha$, one power to convert local energy to energy-at-infinity and one
power to convert the rate measured by local clocks to the rate measured at
infinity. Thus we can immediately define a dissipation functional:
\begin{equation}
D[\phi] \equiv - \f{1}{4 \pi} \int d^2 x \sqrt{\gm} \, \hat{n} \cdot
 \lf \vec{E}_H \times \vec{B}_H \rt \; ,
\end{equation}
where the subscript $H$ denotes that a power of $\alpha$ has been absorbed
to renormalize an otherwise divergent fiducial quantity.

In this manner, we can easily write down the dissipation functional 
for gravity for
which time-differentiating the Hamiltonian is a much more laborious exercise.
The local rate of energy transfer is given by the right-hand side of the heat
transfer equation, Eq. (\ref{heat}). The
Hamiltonian for gravity satisfies two constraint equations with the lapse
and shift vector serving as Lagrange multipliers. Since the membrane 
picture continues to have a gauge freedom associated with time-slicing,
the constraint equation associated with the lapse is not implemented.
This implies that the dissipation is a functional only of the shift. Hence
we have
\begin{equation}
D[v^A] = \int d^2 x \sqrt{\gm} \lf \zeta \theta _H ^2 + 2 \eta 
\sigma _H ^2 + \alpha^2 T^a_b n_a U^b \rt \; ,
\end{equation}
where again the two powers of $\alpha$ have been absorbed to render
finite the quantities with the subscript $H$. Extremizing $D$ with respect
to $v^A$ leads to the membrane equations of motion, enforcing the 
gauge constraint or, equivalently, obeying the principle of minimum
heat production.

\clearpage

\section{Appendix}
In this appendix, we shall prove that Eq. (\ref{zero}) is zero in the limit
that the stretched horizon approaches the true horizon. In that limit, $\alpha
n^a \to l^a$. We shall make liberal use of Gauss' theorem, the
Leibniz rule, and the fact that $h^{ab} n_b = K^{ab} n_b = 0$. 
In order
to use Gauss' theorem, we note that since the ``acceleration''
$a^c \equiv n^d \del _d n^c$ of the normal vector 
(not to be confused with the fiducial acceleration $U^d \del _d U^c$) is zero,
the 4-covariant divergence and the 3-covariant divergence of a vector in
the stretched horizon are equal, Eq. (\ref{div}).

Now, variations in the metric that
are in fact merely gauge transformations can be set to zero. Using a vector
$v^a$ where $v^a$ vanishes on the stretched horizon, 
we can gauge away the variations
in the normal direction so that $\d g_{ab} \to \d h_{ab}$. Then
the left-hand side of Eq. (\ref{zero}) becomes
\begin{eqnarray}
\lefteqn{\int d^3 x \rh h^{bc} \left [ \del _a \lf n^a \d h_{bc} \rt -
\del _c \lf n^a \d h_{ab} \rt \right ] } & & \nonumber \\ 
& = & \int d^3 x \rh \left [ \del _a \lf h^{bc} n^a \d h_{bc} \rt -
\lf \del _a h^{bc} \rt n^a \d h_{bc} 
{}- \del _c \lf h^{bc} n^a \d h_{ab} \rt +
\lf \del _c h^{bc} \rt n^a \d h_{ab} \right ] \nonumber \\
& = &\int d^3 x \rh \left [ \del _a \lf h^{bc} n^a \d h_{bc} \rt 
+ \lf n^c a^b + n^b a^c \rt \d h_{bc}
{}- \lf h^{bc} n^a \d h_{ab} \rt _{|c} \right. \nonumber \\
& & \indent \indent \indent \left. - h^{bc} n^a \d h_{ab} a_c
{}- K n^b n^a \d h_{ab} - a^b n^a \d h_{ab} \right ] \nonumber \\
& & \mbox{(using $h^{bc} = g^{bc} - n^b n^c$, $K_{ab} = + h_a^c \del _c n_b$, 
and $\del_c w^c = w^c_{|c} + w^c a_c$ for $w^c \epsilon {\cal{H}}$)} 
 \nonumber \\
& = & \int d^3 x \rh \left [ \del _a \lf h^{bc} n^a \d h_{bc} \rt 
- K n^b n^a \d h_{ab} \right ] \nonumber \\
& & \mbox{(using Gauss' theorem, and $a^c = 0$)} \nonumber \\
& = & \int d^3 x \rh \left [ \del _a \lf h^{bc} \f {\alpha} 
{\alpha} n^a \d h_{bc} \rt
- K \left [ \d \lf n^b n^a h_{ab} \rt - 
n^a h_{ab} \d n^b - n^b h_{ab} \d n^a \right ] \right ] \nonumber \\
& \to & \int d^3 x \rh \del _a \lf h^{bc} \f {1} {\alpha} l^a \d h_{bc} \rt 
\mbox{(using $h_{ab} n^b = 0$, and $\alpha n^a \to l^a$)} \nonumber \\
& = & \int d^3 x \rh \lf h^{bc} \f {1} {\alpha} l^a \d h_{bc} \rt _{|a} 
\nonumber \\
& = & 0 \; .
\end{eqnarray}

%% file: instanton.tex
\section{Introduction}
A strong hint that the so far classical membrane analogy might remain
valid at a thermodynamic level has already come from the focusing equation, 
Eq. (\ref{energy}), which we saw could be written as the heat transfer
equation, Eq. (\ref{heat}), provided that the temperature was identified
with the surface gravity and the entropy with the area. However, there
are proportionality factors that cannot be determined through classical
arguments alone. In this chapter we shall first show that the 
correct entropy with the correct numerical factor does indeed appear 
from the membrane action, though with a somewhat surprising sign.

One way to interpret the black hole's entropy is as the logarithm of the 
number of modes that propagate along the thin layer 
between the true horizon and the stretched
horizon. The regularity condition, which led to dissipation even in the
classical theory, essentially coarse-grained over these high-frequency modes.
However, it is conceivable that in a quantum theory with benign ultraviolet 
behavior the amount of information contained in that region is finite. 
Einstein gravity is not such a theory but one may still
ask abstractly whether an effective horizon theory could exist at
a quantum level \cite{thooft90,stu}. Quantum effects cause the black
hole to emit radiation. In order to preserve time-evolution unitarity,
one might require the emitted radiation to be correlated with the
interior state of the black hole. 
In this case, the membrane viewpoint remains valid only as a classical
description, since quantum-mechanically the
external universe receives information from the black hole in the form
of deviations of the radiation from thermality; the crucial premise that
the outside universe is emancipated from the internal state of the black
hole is violated. It is important to emphasize, however, that
correlations between the radiation and the horizon itself (as opposed to
the inside of the hole) do not preclude the membrane paradigm. 
Indeed, the fact that the Bekenstein-Hawking
entropy is proportional to the surface area of the black hole suggests that,
even at the quantum level, an effective horizon theory may not be unfeasible.

\section{String Theory or Field Theory?}

In treating the quantum mechanical aspects of black holes, one is faced with
a choice of two different approaches. One of the most exciting recent 
developments in theoretical physics has been the series of spectacular
resolutions that string theory has brought to some longstanding problems 
in black hole physics. In particular, by constructing black hole 
solutions out of collections of D-branes, Strominger and Vafa were
able to provide a microscopic account of black hole entropy in terms of
excitations in higher, compactified dimensions \cite{stromingervafa}.
On the heels of this triumph, came a description of black hole 
radiation as the emission of closed strings from the D-brane 
configuration \cite{callanjuan}. Moreover, this process had an 
underlying unitary theory, indicating that information might not be lost.

These successes and the ultraviolet finiteness of string theory would
indicate that a study of the quantum properties of black holes should
proceed within the framework of string theory and
indeed, in the long run, this may be true. However, at present, there
are several limitations in the string theoretic approach to black holes
that support continuing a field theoretic investigation.
For example, the string calculations are only reliable for very special
black holes (supersymmetric, four or five dimensional, extremal or 
near-extremal). Also the methods are not very general, with the 
entire calculation having to be repeated for each case, though the 
answer for the entropy -- one-fourth the area -- is simple enough.
By contrast, the field theory calculation is short and valid at once
for all black holes in any number of dimensions. 
Similarly, for black hole radiance, string theory has not progressed much
beyond confirming field theory results, despite its unitary promise.
In particular, 
statements about the changing nature of spacetime 
are difficult to make in string theory because the calculations are
not controllable in the regime in which there is a classical spacetime.
By contrast, in the next chapter, we shall obtain a Penrose diagram 
depicting the causal features of the spacetime.
Finally, since the horizon can be in essentially flat space, it is not 
obvious that black hole radiance necessarily
calls for a quantum theory of gravity (although see \cite{schoutensvv}).
For these reasons, we shall take a field theory approach in this thesis.

Before moving on to the calculations, we mention as an aside that there are 
some intriguing parallels between the matrix formulation of M theory
\cite{bfss} and the membrane paradigm: both have a kind of holographic 
principle \cite{thooft93,hologram}, 
both have Galilean equations emerging from a 
Lorentz-invariant theory, and in both the appearance of locality is 
somewhat mysterious. Quite possibly at least some of
these similarities are related to the fact that both are formulated 
on null surfaces.

\section{Entropy}
We have mentioned that black hole entropy was something to be expected,
since the part of the wavefunction or density matrix that lies 
within the black hole is inaccessible and must be traced over.
Thus, the entropy can be thought of as originating in correlations 
across the horizon. 
This entanglement, or geometric, entropy may be computed in field theory 
\cite{srednicki,geometricentropy}. However, the
field theory computations run aground because of uncontrollable
ultraviolet divergences, and so this is not the approach taken here.
Instead, we work with the path integral, making contact with 
thermodynamics by performing an analytic continuation to
imaginary time, $\tau = i t$, so that the path integral 
of the Euclideanized action becomes 
a partition function. For a stationary hole, regularity (or the removal of a 
conical singularity) dictates a period $\beta = \int d \tau = 2 \pi / g_H$ 
in imaginary time \cite{gibbonshawking}, 
where $g_H$ is the surface gravity; for a
Schwarzschild hole, $\beta = 8 \pi M$. 
This is the inverse Hawking temperature in
units where $\hbar = c = G = k_B = 1$. 
The partition function is then the path integral over all Euclidean metrics
which are periodic with period $2 \pi / g_H$ in imaginary time. 
We can now evaluate the partition function in a stationary phase approximation:
\begin{equation}
Z = \int D g_E^{ab} \exp \lf - \f{1}{\hbar} \lf S^E_{\rm out}[g_E^{ab}] 
+ S^E_{\rm surf}[h_E^{ab}] \rt \rt \approx \exp \lf - \f{1}{\hbar}
\lf S^E_{\rm out}[g^{ab}_{E \, {\rm cl}}] + S^E_{\rm surf}[h^{ab}_{E \, {\rm cl}}] \rt \rt .
\end{equation}

The external action itself can be written as $S_{\rm out} = S_{\rm bulk} 
+ S_{\infty}$, where $S_{\rm bulk}$ is zero for a black 
hole alone in the universe. The boundary term $S_{\infty}$ is the integral
of the extrinsic curvature of the boundary of spacetime. In fact, a term
proportional to the surface area at infinity can be included in $S_{\infty}$
without affecting the Einstein equations since the metric is held fixed at
infinity during variation. In particular, the proportionality 
constant can be chosen so that the action for all of spacetime is 
zero for Minkowski space:
\begin{equation}
S_{\infty} = \f {1}{8 \pi} \int d^3 x \rh [K] \; ,
\end{equation}
where $[K]$ is the difference in the trace of the extrinsic curvature at
the spacetime boundary for the metric $g_{ab}$ and the flat-space metric $\eta 
_{ab}$. With this choice, the path integral has a properly normalized
probabilistic interpretation. The Euclideanized value of $S_{\infty}$ for
the Schwarzschild solution is then \cite{gibbonshawking}
\begin{equation}
S^E_{\infty} = \lim_{r\to\infty} \f{1}{8 \pi} \lf - 32 \pi ^2 M \rt 
\left [ \lf 2r - 3M \rt -2r \lf 1 - \f{2M}{r} \rt ^{1 / 2} \right ] 
= + 4 \pi M^2 \; .
\label{Sbound}
\end{equation}

To obtain an explicit action for the membrane, we must integrate
its variation, Eq. (\ref{cancel}):
\begin{equation}
\d S_{\rm surf}[h^{ab}] = - \f {1}{16 \pi} \int d^3 x \rh \lf K h_{ab} - K_{ab}
\rt \d h^{ab} \; .
\end{equation}
We see that
\begin{equation}
S_{\rm surf}[h^{ab}] = \int d^3 x \rh \lf B_{ab} h^{ab} - b \rt
\end{equation}
is a solution, provided that the (undifferentiated) source terms are 
$B_{ab} = (+ 1 / 16 \pi) K_{ab}$ and $b = (- 1 / 16 \pi) K$. 
This action has the form of surface matter plus a
negative cosmological constant in three dimensions. 
The value of the membrane action 
for a solution to the classical field equations is then
\begin{equation}
S_{\rm surf}[h^{ab}_{\rm cl}] = + \f {1} {8 \pi} \int d^3 x \sqrt{-h_{\rm cl}} 
K_{\rm cl} 	\; . \label{Ssurf}
\end{equation}

To evaluate this, we can take our fiducial world-lines $U^a$ to be 
normal to the \mbox{isometric} time-slices of constant Schwarzschild time. 
The stretched horizon is then a surface of
constant Schwarzschild $r$. Hence $\alpha = \lf 1 - 2M / r \rt ^{1 / 2}$,
$\theta = 0$, and $K = g + \theta = g$, the unrenormalized surface gravity
of the stretched horizon. Inserting these into Eq. (\ref{Ssurf}), we 
find that the Euclidean action is
\begin{equation}
S^E_{\rm surf} = \lim_{r\to r_H} \f{1}{8 \pi} \lf \int - d \tau \rt 
\alpha 4 \pi r^2 g = - \pi {r_H}^2 = - 4 \pi M^2 \; ,
\end{equation}
where $r_H = 2M$ is the black hole's radius, and $g_H = \alpha g = 1 / 4M$ 
is its renormalized surface gravity.

The Euclidean membrane action exactly cancels the external action, Eq.
(\ref{Sbound}). Hence the entropy is zero! 
That, however, is precisely what makes the
membrane paradigm attractive: to an external observer, 
there is no black hole -- only a membrane -- and so neither a 
generalized entropy nor a strictly obeyed second law of thermodynamics. 
The entropy of the
outside is simply the logarithm of the number of quantum states of the
matter outside the membrane. This number decreases as matter leaves the
external system to fall through and be dissipated by the membrane. When all
matter has fallen into the membrane, the outside is in a single state -- vacuum
{}-- and has zero entropy, as above.

To recover the Bekenstein-Hawking entropy, we must then use not the
combination of external and membrane actions, which gave the entropy of the
external system, but the combination of the {\em internal} and 
membrane actions,
\begin{equation}
Z_{B-H} = \int D g_E^{ab} \exp \lf - \f{1}{\hbar}\lf S^E_{\rm in}[g_E^{ab}] -
S^E_{\rm surf}[h_E^{ab}] \rt \rt \; ,
\end{equation}
where now $S_{\rm surf}$ is subtracted [see Eq. (\ref{split})]. 
With $S_{\rm in} = \int d^4 x \rg R = 0$, 
the partition function for a Schwarzschild hole in
the stationary phase approximation is
\begin{equation}
Z_{B-H} \approx \exp \lf - \f{1}{\hbar} \lf + 4 \pi M^2 \rt \rt \; ,
\end{equation}
from which the Bekenstein-Hawking entropy, $S_{B-H}$, immediately follows:
\begin{equation}
S_{B-H} = \beta \lf M + \f{\ln Z_{B-H}}{\beta} \rt 
= 8 \pi M \lf M - \f{1}{8 \pi M} 4 \pi M^2 \rt = \f{1}{4} A \; ,
\end{equation}
which is the celebrated result.

For more general stationary (Kerr-Newman) holes, the Helmholtz free energy
contains additional ``chemical potential'' terms corresponding to the
other conserved quantities, $Q$ and $J$,

\begin{equation}
F = M - TS - \Phi Q - \Omega J \; ,	\label{free}
\end{equation}
where $\Phi = Q / r_+$ and $\Omega = J / M$, where $r_+$ is the
Boyer-Lindquist radial coordinate at the horizon. For a charged hole,
the action also contains electromagnetic terms. The surface electromagnetic
term, Eq. (\ref{charge}), has 
the value $(1 / 4 \pi) \int d^3 x \rh F^{ab} A_a n_b$.
However, in order to have a regular vector potential, we must gauge transform
it to $A_a = \lf Q / r - \Phi \rt \del _a t$ which vanishes on the surface.
Hence, the surface action is again given by the gravitational term, which
has the Euclideanized value $S^E_{\rm surf} = - \pi r^2_+$. It is easy to
verify using Eq. (\ref{free}) that this again leads to a black hole entropy 
equal to one-fourth of the horizon surface area and an external entropy
of zero.

For nonstationary black holes, the extrinsic
curvature also includes a term for the expansion of the horizon, $K = g +
\theta$. Inserting this into the surface action enables us to calculate the
instantaneous entropy as matter falls into the membrane in a nonequilibrium
process. Of course, like the horizon itself, the entropy grows acausally.

\section{Pictures of Hawking Radiation}
We now turn to the phenomenon of black hole radiance. Although
several derivations of Hawking radiation exist in the literature 
\cite{swh,gibbonshawking},
none of them correspond very directly to either of the two
heuristic pictures that are most commonly proposed as ways to visualize
the source of the radiation. According to one picture, the
radiation arises by a process similar to Schwinger electron-positron pair
creation in a constant electric field.  The idea is that the energy of
a particle changes sign as it crosses the horizon, so that a
pair created just inside or just outside the horizon can
materialize with zero total energy, after one member of the pair has
tunneled to the opposite side.  
In the second picture, we work with the effective membrane 
representation of the horizon.
Hawking radiation is then a special property of the membrane: its 
tendency towards spontaneous emission, as if it had a nonzero temperature. 

Here we shall show that either of these pictures can in fact 
be used to provide short, direct semi-classical derivations of black hole
radiation.  In both cases energy conservation plays a fundamental
role; one must make a transition between states with the same total
energy, and the mass of the residual hole must go down as it radiates.
Indeed, it is precisely the
possibility of lowering the black hole mass which ultimately drives the
dynamics. This supports the idea that, in quantum gravity, 
black holes are properly regarded as highly excited states.

In the standard
calculation of Hawking radiation the background geometry is considered
fixed, and energy conservation is not enforced. (The geometry is not
truly static, despite appearances, 
as there is no global Killing vector.)  Because we are treating
this aspect more realistically, we must -- and do -- find corrections
to the standard results. These become quantitively significant when
the quantum of radiation carries a substantial fraction of the 
mass of the hole.

\section{Tunneling}

To describe across-horizon phenomena, it is necessary
to choose coordinates which, unlike Schwarzschild coordinates, are not
singular at the horizon.  A particularly suitable choice is obtained by 
introducing a time coordinate,
\begin{equation}
t = t_s + 2 \sqrt {2 M r} + 2M \ln \f {\sqrt{r} - \sqrt{2M}}
{\sqrt{r} + \sqrt{2M}} \; ,
\end{equation}
where $t_s$ is Schwarzschild time.  With this choice, the line element reads
\begin{equation}
ds^2 = - \lf 1 - \f{2M}{r} \rt dt^2 + 2 \, \sqrt{\f{2M}{r}} \,
 dt \, dr + dr^2 + r^2 d \Omega^2 \; .	\label{ds2}
\end{equation}
There is now no singularity at $r= 2M$, and the true character
of the spacetime, as being stationary but not static, is manifest. 
These coordinates were first introduced by the French mathematician 
Paul Painlev\'e \cite{painleve} and the Swedish opthalmologist and Nobel
laureate Allvar Gullstrand \cite{gullstrand}, who used them to criticize 
general relativity for allowing singularities to come and go! Their utility
for studies of black hole quantum mechanics was emphasized more recently in
\cite{line}.

For our purposes, the crucial features of these coordinates are that they
are stationary and 
nonsingular through the horizon.  Thus it is possible to define an effective 
``vacuum'' state of a quantum field by requiring that it annihilate
modes which carry negative frequency with respect to $t$; such a
state will look essentially empty (in any case, nonsingular) 
to a freely-falling observer as he or she
passes through the horizon. This vacuum differs strictly from the
standard Unruh vacuum, defined by requiring positive frequency with
respect to the Kruskal coordinate $U = - \sqrt{r - 2M} \exp \lf -{t_s - r
\over 4 M} \rt$ \cite{unruh}. The difference, however, shows up only in
transients, and does not affect the late-time radiation.  

The radial null geodesics are given by
\begin{equation}
\dot{r} \equiv \f{dr}{dt} = \pm 1 - \sqrt{2M \over r} \; , \label{null}
\end{equation}
with the upper (lower) sign in Eq. (\ref{null})
corresponding to outgoing (ingoing) geodesics, under the implicit 
assumption that $t$ increases towards the future. These equations are modified 
when the particle's self-gravitation is taken
into account. Self-gravitating shells
in Hamiltonian gravity were studied by Kraus and Wilczek \cite{per}. They 
found that, when the black hole mass is held fixed and the total ADM mass
allowed to vary, a shell of energy $\w$ moves in the geodesics of a spacetime
with $M$ replaced by $M+\w$. If instead
we fix the total mass and allow the hole mass to fluctuate, 
then the shell of energy $\w$ travels on the 
geodesics given by the line element
\begin{equation}
ds^2 = - \lf 1 - {2 ( M - \w ) \over r} \rt dt^2 
+ 2 \, \sqrt{ {2 (M - \w ) \over r}} \, dt \, dr + dr^2 + r^2 d \Omega^2 \; ,
\end{equation}
so we should use Eq. (\ref{null}) with $M \to M - \w$.

Now one might worry that, since the typical wavelength of the radiation
is of the order of the size of the black hole, 
a point particle description might be
inappropriate. However, when the outgoing wave is traced back towards the
horizon, its wavelength, as measured by local fiducial observers, is
ever-increasingly blue-shifted. Near the horizon, the radial wavenumber
approaches infinity and the point particle, or WKB,
approximation becomes in fact excellent.

The imaginary part of the action for an s-wave
outgoing positive energy particle  which crosses the horizon
outwards from $r_{\rm in}$ to $r_{\rm out}$ can be expressed as
\begin{equation}
{\rm Im}~ S = {\rm Im} \int_{r_{\rm in}}^{r_{\rm out}} p_r \, dr = 
{\rm Im} \int_{r_{\rm in}}^{r_{\rm out}} \! \!
\int_0^{p_r} d p'_r \, dr \; . \label{integrals}
\end{equation} 
Remarkably, this can be evaluated without entering into the details of
the solution, as follows.  We multiply and divide the integrand by the
two sides of Hamilton's equation
$\dot{r} = +\left. {dH \over d p_r} \right | _r \,$,  
change variable from momentum to
energy, and switch the order of integration to obtain
\begin{equation}
{\rm Im}~ S = {\rm Im} \int_0^{+\w} \!\!
\int_{r_{\rm in}}^{r_{\rm out}} {dr \over 1 - \sqrt {2 \lf M -
\w' \rt \over r}} \, \lf - d \w' \rt \; ,
\end{equation} 
where the minus sign appears because $H = M - \w'$. 
But now the integral can be done by deforming the contour, so as to
ensure that positive energy solutions decay in time (that is, into the
lower half $\w'$ plane). In this way we obtain
\begin{equation}
{\rm Im}~ S = + 4 \pi \w \lf M - {\w \over 2} \rt \; , \label{answer}
\end{equation} 
provided $r_{\rm in} > r_{\rm out}$. To understand this ordering --
which supplies the correct sign -- we observe that 
when the integrals in Eq. (\ref{integrals})
are not interchanged, and with the contour evaluated via the 
prescription $\w \to \w - i \eps$, we have
\begin{equation}
{\rm Im}~ S = + {\rm Im} \int_{r_{\rm in}}^{r_{\rm out}} \!\! 
\int_M^{M - \w} {d M' \over 1 - \sqrt{{2 M' \over r}}} \, dr
= {\rm Im} \int_{r_{\rm in}}^{r_{\rm out}} - \pi r \, dr \; .
\end{equation}
Hence $r_{\rm in} = 2 M$ and $r_{\rm out} = 2 \lf M - \w \rt$. 
(Incidentally, comparing the above equation with Eq. (\ref{integrals}), 
we also find that ${\rm Im}~ p_r = - \pi r$.) Thus, over the course of 
the classically forbidden trajectory, the outgoing particle travels 
radially inward with the apparent horizon to
materialize at the {\em final} location of the horizon, 
viz. $r = 2 \lf M - \w \rt$.

Alternatively, and along the same lines, Hawking
radiation can also be regarded as pair creation {\em outside\/} the horizon,
with the negative energy particle tunneling into the black hole. Since such
a particle propagates backwards in time, we have to reverse time in the
equations of motion. From the line element, Eq. (\ref{ds2}), we see that
time-reversal corresponds to 
$\sqrt{\f{2M}{r}} \to - \sqrt{\f{2M}{r}}$. Also, since the anti-particle
sees a geometry of fixed black hole mass, the upshot of self-gravitation
is to replace $M$ by $M + \w$, rather than $M - \w$.
Thus an ingoing negative energy particle has
\begin{equation}
{\rm Im}~ S = {\rm Im} \int _{0} ^{-\w} \!\! \int _{r_{\rm out}} ^{r_{\rm in}} 
{dr \over -1 + \sqrt{{2 \lf M + \w' \rt \over r}} } \, d \w' 
= + 4 \pi \w \lf M - \f {\w}{2} \rt \; ,
\end{equation}
where to obtain the last equation we have used Feynman's ``hole theory'' 
deformation of the contour: $\w' \to \w' + i \eps$.

Both channels -- particle or anti-particle tunneling -- 
contribute to the rate for the Hawking process so, 
in a more detailed calculation, one would have to add their
amplitudes before squaring in order to obtain the
semi-classical tunneling rate. That, however, only affects the pre-factor.
In either treatment, the exponential part of the semi-classical emission rate, 
in agreement with \cite{k5}, is
\begin{equation}
\Gamma \sim e^{-2 \, {\rm Im}~ S} = e^{-8 \pi \w \Mw2 } = e^{+ \Delta 
S_{\rm B-H}} \; , \label{rate}
\end{equation}
where we have expressed the result more naturally in terms of the change 
in the hole's Bekenstein-Hawking entropy, $S_{\rm B-H}$.
When the quadratic term is neglected, Eq. (\ref{rate})
reduces to a Boltzmann factor for a particle with energy $\w$ at
the inverse Hawking temperature $8 \pi M$.  The $\w^2$ correction
arises from the physics of energy conservation, which (roughly
speaking) self-consistently raises the effective temperature of the
hole as it radiates.  
That the exact result must be correct can be seen on physical grounds by 
considering the 
limit in which the emitted particle carries away the entire mass and 
charge of the black hole (corresponding to the transmutation of the
black hole into an outgoing shell). There can be only one such outgoing 
state. On the other hand, there are $\exp \lf S_{\rm B-H} \rt$ states in total.
Statistical mechanics then asserts that the probability of finding a
shell containing all the mass of the black hole is proportional
to $\exp \lf - S_{\rm B-H} \rt$, as above.

Following standard arguments, Eq. (\ref{rate}) with the quadratic term
neglected implies
the Planck spectral flux appropriate to an inverse temperature of $8 \pi M$:
\begin{equation}
\rho \lf \omega \rt = \f{d \w}{2 \pi} \f{| \, T \lf \omega \rt | ^2} 
{e^{+ 8 \pi M \omega } - 1} \; ,
\end{equation}
where $| \, T \lf \omega \rt | ^2$ is the frequency-dependent (greybody) 
transmission co-efficient for the
outgoing particle to reach future infinity without back-scattering.
It arises from a more complete treatment of the modes, whose
semi-classical behavior near the turning point we have been discussing.

\subsection{Tunneling from a Charged Black Hole}

When the outgoing radiation carries away the black hole's charge, 
the calculations are complicated by the fact that the trajectories 
are now also subject to electromagnetic forces. 
Here we restrict ourselves to uncharged radiation coming from a 
Reissner-Nordstr\"om black hole. The calculation then proceeds in an 
exactly similar fashion as in the case of Schwarzschild holes but,
for completeness, we shall run through the corresponding equations.

The charged counterpart to the Painlev\'e line element is
\begin{equation}
ds^2 = - \lf 1 - {2M \over r} + {Q^2 \over r^2} \rt dt^2 + 2 \, 
\sqrt{{2M \over r} - {Q^2 \over r^2}} \,  dt \, dr + 
dr^2 + r^2 d \Omega^2 \; ,
\end{equation}
which is obtained from the standard line element by 
the rather unedifying coordinate transformation,
\begin{eqnarray}
t & = & t_r + 2 \Root +M \ln \lf {r - \Root \over r + \Root} \rt \nonumber \\
& & \indent \indent + {Q^2 - M^2 \over \root} \, 
{\rm arctanh} \lf {\root \Root \over Mr} \rt \; ,
\end{eqnarray}
where $t_r$ is the Reissner time coordinate. The equation of motion
for an outgoing massless particle is
\begin{equation}
\dot{r} \equiv \f{dr}{dt} = + 1 - \sqrt{{2M \over r} - {Q^2 \over r^2}} \; ,
\end{equation}
with $M \to M - \w$ when self-gravitation is included. The imaginary part
of the action for a positive energy outgoing particle is
\begin{equation}
{\rm Im}~ S = {\rm Im} \int_0^{+\w} \!\!
\int_{r_{\rm in}}^{r_{\rm out}} {dr \over 1 - 
\sqrt {{2 \lf M - \w' \rt \over r} - {Q^2 \over r^2}} } \, \lf - d \w' \rt \; ,
\end{equation} 
which is again evaluated by deforming the contour in accordance with
Feynman's $w' \to w' - i \eps$ prescription. The residue at the pole
can be read off by substituting $u \equiv \sqrt{2 \lf M - \w' \rt r - Q^2}$. 
Finally, the emission rate is
\begin{equation}
\Gamma \sim e^{-2 \, {\rm Im}~ S} 
= e^{- 4 \pi \lf 2 \w \Mw2 - (M- \w) \sqrt{(M-\w)^2 - Q^2} + M \root \rt}
= e^{+ \Delta S_{\rm B-H}} \; .
\end{equation}
To first order in $\w$, this is consistent with Hawking's result of
thermal emission at the Hawking temperature, $T_H$, for a charged black hole:
\begin{equation}
T_H = {1 \over 2 \pi} {\root \over \lf M + \root \rt ^2} \; .
\end{equation}
But again, energy conservation implies that the exact result has
corrections of higher order in $\w$; these can all be collected to 
express the emission rate as the exponent of the change in entropy.

We conclude this section by noting that only local physics has gone into
our derivations. There was neither an appeal to Euclidean gravity nor a
need to invoke an explicit collapse phase. 
The time asymmetry leading to outgoing
radiation arose instead from use of the ``normal'' local contour deformation
prescription in terms of the nonstatic coordinate $t$.

\section{Emissions from the Membrane}

As we have seen in the previous chapter, an outside observer can adopt 
another, rather different, description of a black hole, regarding
the horizon effectively as a membrane at an inner boundary of
spacetime. But again, quantum uncertainty in the position of the membrane
complicates the issue in an essential way.

Before presenting the calculation, we briefly outline the strategy here.
Starting with a fundamental action for the bulk, we obtain the membrane
action as well as the bulk equations of motion. After continuing the equations
to complex time, we look for solutions that connect the Lorentzian geometries
before and after emission. Finally, we evaluate the action for our instanton
to obtain the semi-classical rate. 

The external action for an uncharged massless scalar field minimally 
coupled to gravity is
\begin{equation}
S [\phi, g_{ab}] = + \f{1}{16 \pi} \int d^4 x \rg R - 
{1 \over 2} \int d^4 x \rg \lf \pl \phi \rt ^2 + S_{\pl M} + S_{\rm surf} \; .
\end{equation}
Here the bulk terms have support only outside the black hole, and $S_{\pl M}$ 
is the term at the external boundaries of the spacetime manifold that is
needed to obtain Einstein's equations \cite{gibbonshawking}. The 
membrane action takes the classical value,
\begin{equation}
S_{\rm surf}[\phi, h_{ab}] = + \f{1}{8 \pi} \int d^3 x \rh K - 
\int d^3 x \rh \phi J_s \; ,	\label{Smemb}
\end{equation}
where $h_{ab}$ is the metric induced on the stretched horizon, $h$ is its
determinant, and $K \equiv + \del_a n^a$ is the trace of the membrane's
extrinsic curvature. The scalar field source, $J_s$, 
induced on the membrane is
\begin{equation}
J_s = + n^a \del _a \phi \; , 	\label{jscalar}
\end{equation}
with $n^a$ the outward-pointing space-like normal to the membrane.

The field equations are Einstein's equations and the source-free 
Klein-Gordon equation. The energy-momentum tensor is
\begin{equation}
T_{ab} = \del_a \phi \del_b \phi - \f{1}{2} g_{ab} \del_c \phi \del^c \phi \; ,
\end{equation}
whose trace is simply
\begin{equation}
T = - \lf \del \phi \rt ^2 \; .	\label{trace}
\end{equation}
{}From this we see that the Einstein-Hilbert and Klein-Gordon bulk
actions cancel, so surface actions account for all the classical and
semi-classical physics.

Next, we seek an instanton solution that
connects the Lorentzian Schwarzschild geometry
of mass $M$ with a horizon at $r_H = 2M$, to a similar geometry of
mass $M - \w$ and a horizon at $r_H = 2 \lf M - \w \rt$. 
At this point, it is important to distinguish the stretched horizon, a 
surrogate for the globally-defined and acausal true horizon, 
from the locally-defined apparent horizon. For an evaporating hole, the 
true horizon (hence the stretched horizon) 
lies {\it inside} the apparent horizon; this is because the acausal true
horizon shrinks in anticipation of future emissions before the local geometry
actually changes. Hence, our analytically-continued solution must 
describe the geometry {\it interior} to the apparent horizon.

Now, in the usual analytic continuation ($t = - i \tau$), 
the apparent horizon is at the origin, and
$r < 2M$ is absent from the Euclidean section of the geometry. 
However, there is no real need for a Euclidean section. Euclidean solutions
may have positive Euclidean action, but the (Lorentzian) action for a general 
tunneling motion in a time-dependent setting need not be purely
imaginary, and the instanton can be a more complicated trajectory in 
the complex time plane. It should not be surprising then that for a
time-dependent shrinking black hole, one has to consider 
intermediate metrics of arbitrary signature. Indeed,
the usual analytic continuation prescription yields, for $r < 2M$,
\begin{equation}
ds ^2 = - \lf \f{2M}{r} - 1 \rt d \tau ^2 - {\lf \f{2M}{r} -1 \rt}^{-1} {dr^2}
+ r^2 d \Omega ^2	\label{interior}
\end{equation}
which has two time-like coordinates. Substituting $x \equiv 4M \lf \f{2M}{r}
- 1 \rt ^{1/2}$, we eliminate the \mbox{coordinate} 
singularity at $r = 2M$
to obtain a spacetime with topology $S^1 \times S^2 \times R$, in which
$\tau$ has a period of $8 \pi M$ about the apparent horizon. The line element
now describes the complexified geometry interior to the apparent horizon;
moreover, it is regular everywhere except at the real singularity at 
$r = 0$.

In addition, we note that Birkhoff's theorem -- spherically symmetric vacuum
solutions are stationary -- is valid irrespective of
the metric signature. Hence, one Schwarzschild solution can go to another
only if there is matter present. The matter is produced by the membrane
during the complex time process, and materializes as the shell. We expect the
membrane to be able to produce matter since it is also able to destroy
(dissipate) matter; Hawking emission is the counterpart of absorption.

As matter is emitted during the emission process, the membrane
is subject to a changing interior geometry which in turn implies a changing
periodicity. In order to adapt Eq. (\ref{interior}) to a geometry with 
changing periodicity, we guess that the line element takes the form
\begin{equation}
ds^2 = - u(\psi) x^2 d \psi^2 - v(\psi, x) dx^2 + r^2 (\psi, x) d \Omega ^2
\end{equation}
which has a radial ($x$) and an angular ($\psi$) coordinate, both with
dimensions of length. The period changes if $u(\psi)$, and $v(\psi, x)$ are 
not constant functions of $\psi$. The shell and the apparent horizon 
are at $x=0$; the shell trajectory is parametrized by $\psi$. 
A convenient choice of $\psi$ is one for which the shell's radius 
decreases linearly with $\psi$. Then the radius of the apparent horizon is
\begin{equation}
r_H = 2 ( M - b \psi ) \; , \label{radius}
\end{equation}
where $b$ is dimensionless.

The important Christoffel symbols are
\begin{equation}
\Gamma^{\psi}_{\psi \psi} = + {\dot{u} \over 2u} \; , \; \; 
\Gamma^{\psi}_{xx} = - {\dot{v} \over 2ux^2} \; , \; \;
\Gamma^{\psi}_{\psi x} = {1 \over x}
\end{equation}
\begin{equation}
\Gamma^{x}_{\psi \psi} = - {ux \over v} \; , \; \;
\Gamma^{x}_{xx} = + {v' \over 2v} \; , \; \;
\Gamma^{x}_{\psi x} = {\dot{v} \over 2v} \; ,
\end{equation}
where $\dot{}$ and ${}'$ denote differentiation by $\psi$ and $x$, 
respectively. A normalized trajectory of constant $x$ has
\begin{equation}
U^2 = - 1 \; , \; \; U^{\psi} = {1 \over x \sqrt{u}} \; ,
\end{equation}
so its proper acceleration has magnitude
\begin{equation}
\alpha = {1 \over x \sqrt{v}} \; .	\label{magacc}
\end{equation}
The normal vector, $n^a$, normal to $U^a$ obeys
\begin{equation}
n^2 = -1 \; , \; \; n^{x} = \f{1}{\sqrt{v}} \; .
\end{equation}
Hence the trace of the extrinsic curvature of a surface of constant $x$ is
\begin{equation}
K = \del_a n^a = \f{1}{\sqrt{v}} \lf {1 \over x}+{2 \over r} r' \rt \; .
\end{equation}
We will also need one component of the Ricci tensor:
\begin{equation}
R_{\psi x} = {1 \over r} \lf - 2 \dot{r}' +
{2 \over x} \, \dot{r} + r' {\dot{v} \over v} \rt \; . 	\label{ricci}
\end{equation}
Now the flux of energy-at-infinity (per local proper time), ${\cal F}$, is
\begin{equation}
{\cal F} = {1 \over 4 \pi r^2} {d \w' \over d \tau} = 
- {1 \over 4 \pi r^2} {\dot{r} \over 2} {1 \over x \sqrt{u}} \; .
\end{equation}
This is related to the local stress tensor by 
${\cal F} = - \sqrt{u} x T_{ab} U^{a} n^{b}$. Hence
\begin{equation}
T_{\psi x} = {1 \over 2} {1 \over 4 \pi r^2} \sqrt{{v \over u}}
{\dot{r} \over x} \; . \label{Stress}
\end{equation}

Comparing Eq. (\ref{Stress}) with Eq. (\ref{ricci}) in the $x \to 0$ limit, 
we find
\begin{equation}
\sqrt{{u \over v}} = {1 \over 2r} \; .	\label{uv}
\end{equation}
Incidentally, by Eq. (\ref{magacc}), this implies that 
the ``temperature-at-infinity'' is
\begin{equation}
T_{\infty} = x \sqrt{u} {\alpha \over 2 \pi} = {1 \over 4 \pi r_H} \; ,
\end{equation}
indicating a changing periodicity; as $r_H$ decreases from $2M$ to $2(M - \w)$,
the temperature varies accordingly.

Now, for single-particle emission, both the true and the apparent horizon
start out at $r = 2M$ and finish at $r = 2(M-\w)$. Thus, the membrane moves 
along the apparent horizon. But now this is mapped
to the origin by Euclideanization, so the stretched horizon
has a vanishing proper volume element,
even in Euclidean space. Thus, in the absence of a compensating divergence
in the integrand, membrane integrals are zero. In particular, the scalar
field current induced on the membrane has no divergence so the scalar
part of the membrane action vanishes. Therefore the entire 
contribution to the emission rate comes from the gravitational term.

To evaluate that, note that there is a factor of $x$ in the action, contained
in $\sqrt{-g_{\psi \psi}}$. Then, as $x \to 0$,
\begin{equation}
x K \to {1 \over \sqrt{v}} \; .
\end{equation}
Combining this with Eqs. (\ref{radius}) and (\ref{uv}), we have
\begin{equation}
S = {1 \over 8 \pi} \int d^3 x \rh K = 
- {i \over 2} \int r^2 \sqrt{{u \over v}} d \psi 
= - {i \over 16 b} \Delta r^2 \; .
\end{equation}
When $b = {1 \over 8 \pi}$, we obtain the desired result, Eq. (\ref{answer}).
This may be fixed by matching it with the rate $\exp (- 4 \pi M )$ for
emission of the entire mass of the hole. Alternatively, we note that for
single-particle emission from a Schwarzschild hole, there is no scale other
than the mass of the hole. Since the instanton simply scales the 
horizon radius, we must have $K_H = g_H = {1 \over 4M}$ throughout the motion.
Hence the proper length along the stretched horizon, $d \tau \equiv 
\sqrt{u} d \psi$, must also scale. Thus, we have
\begin{equation}
\beta = \int d \tau = 8 \pi M \Rightarrow d \tau = 8 \pi d M \; ,
\end{equation}
so that $b = {1 \over 8 \pi}$, as required.
We note here that $b$ can be written as
${M \over \beta}$ where $\beta$ is the inverse temperature. As we shall 
see, with this form as an ansatz, the membrane action gives the
right rate for emission from Reissner holes.

\subsection{Emission from a Charged Membrane}

For emission from charged black holes, there are a few
modifications to the preceding equations. 
The action now has an additional term because of the
electromagnetic field:
\begin{equation}
S [g_{ab}, A_a, \phi] = + \f{1}{16 \pi} \int d^4 x \rg R - 
{1 \over 2} \int d^4 x \rg \lf \pl \phi \rt ^2 - \f{1}{16 \pi} \int d^4 x 
\rg F^2 + S_{\rm surf} \; ,
\end{equation}
which yields a membrane action of
\begin{equation}
S_{\rm surf}[h_{ab}, A_a, \phi] = + \f{1}{8 \pi} \int d^3 x \rh K - 
\int d^3 x \rh \phi J_s + \int d^3 x \rh j_s^a A_a \; .
\end{equation}
In addition to the scalar term induced on the membrane, there is now also
an electromagnetic current, Eq. (\ref{jmaxwell}), 
\begin{equation}
j_s^a = + {1 \over 4 \pi} F^{ab} n_b \; ,
\end{equation}
as we saw in the previous chapter.
The stress tensor is
\begin{equation}
T_{ab} = \del _a \phi \del _b \phi - \f{1}{2} g_{ab} \del_c \phi \del ^c \phi
+ {1 \over 4 \pi} \lf F_{ac} F_b^{\, c} - {1 \over 4} g_{ab} F^2 \rt \; ,
\end{equation}
but the trace is still given by Eq. (\ref{trace}), so the Einstein-Hilbert
and Klein-Gordon bulk actions again cancel.
Now, in the absence of sources, the Maxwell action
can be expressed as a surface integral:
\begin{equation}
- \f{1}{16 \pi} \int d^4 x \rg F^2 
= +\f{1}{8 \pi} \oint d^3 x \rh n_a F^{ab} A_b \; ,
\end{equation}
where the sign on the right-hand side stems from choosing the normal to point
into the bulk. This term is added to the membrane term to give a total
electromagnetic action of
\begin{equation}
S^{\rm EM}_{\rm surf} = + \f{1}{8 \pi} \int d^3 x \rh F^{ab} A_a n_b \; .
\end{equation}
Thus, we have succeeded in eliminating all the 
bulk terms in the action. To evaluate the surface terms, we note first of
all that, because the volume element along the membrane vanishes, the 
scalar membrane term contributes nothing to the rate.
The electromagnetic and gravitational membrane terms can be combined, in
the $x \to 0$ limit, to give
\begin{equation}
S_{\rm surf} = - {i \over 8 \pi} \int d \psi \, 4 \pi r^2 \left [ 
{\sqrt{M^2 - Q^2} \over r^2} + {Q^2 \over r^3} \right ] \; ,
\end{equation}
where we have skipped the steps analogous to Eqs. (\ref{radius}) -
(\ref{uv}). Using 
\begin{equation}
r_H = M - b \psi + \sqrt{(M- b \psi)^2 - Q^2} \; , 
\end{equation}
and our ansatz,
\begin{equation}
b \equiv {M \over \beta} = {M \over 2 \pi} {\sqrt{M^2 -Q^2} \over r^2} \; ,
\end{equation}
we have that
\begin{equation}
S_{\rm surf} = -i {\pi \over 2} \Delta r_+^2 \; ,
\end{equation}
which yields the correct tunneling rate.

When the hole emits charged radiation, the analysis become more
complicated. However, one can consider the case in which the emitted
radiation has the same charge-to-mass ratio as the hole itself. Then
the problem again becomes one of scaling. Letting $Q \equiv \eta M$
with $|\eta| < 1$, we have
\begin{equation}
\beta = 2 \pi {\lf 1 + \sqrt{ 1 - \eta ^2} \rt ^2 \over
\sqrt{1 - \eta ^2}} M \Rightarrow d \tau = 2 \pi 
{\lf 1 + \sqrt{ 1 - \eta ^2} \rt ^2 \over \sqrt{1 - \eta ^2}} d M \; .
\end{equation}
With $K_H = g_H$, we have
\begin{equation}
S = -i \int \pi M \lf 1 + \sqrt{1 - \eta ^2} \rt ^2 dM 
= -i {\pi \over 2} \Delta r_+^2 \; .
\end{equation}
Calling the change in the hole's charge $q$, the emission rate
to first order in $\w$ and $q$ is Hawking's result,
\begin{equation} 
\Gamma \sim e^{-{2 \pi \over g_H} ( \w - q \Phi)} \; ,
\end{equation}
where $\Phi \equiv + Q / r_+$ is the electromagnetic scalar potential
at the horizon; emissions which discharge the membrane are favored.
The exact rate is again proportional to the exponent of the change
in entropy.

%% file: vaidya.tex
\section{Introduction}
It is challenging to envision a plausible global structure for a spacetime
containing a decaying black hole. If information is not lost in 
the process of black hole decay, then the final
state must be uniquely determined by the initial state, and vice versa.
Thus a post-evaporation space-like hypersurface must lie within the future
domain of dependence of a pre-evaporation Cauchy surface. One would like to
have models with this property that support approximate (apparent) horizons.

In addition, within the framework of general relativity,
one expects that singularities will form inside black holes \cite{singularity}.
If the singularities are time-like, one can
imagine that they will go over into the world-lines of additional degrees
of freedom occurring in a quantum theory of gravity.
Ignorance of the nature of these degrees of freedom is reflected in
the need to apply boundary conditions at such singularities.
(On the other hand, boundary conditions at future space-like singularities
represent constraints on the initial conditions; it is not obvious how a
more complete dynamical theory could replace them with something more
natural.)

In this chapter,
we use the charged Vaidya metric to obtain a candidate macroscopic 
Penrose diagram for the
formation and subsequent evaporation of a charged black hole,
thereby illustrating how predictability might be retained. We do this
by first extending the charged Vaidya metric past its coordinate
singularities, and then joining together patches of spacetime that
describe different stages of the evolution.

\section{Extending the Charged Vaidya Metric}

The Vaidya metric \cite{vaidya} and its charged generalization
\cite{plebanskistachel,bonnorvaidya}
describe the spacetime geometry of unpolarized radiation, represented
by a null fluid, emerging from a spherically symmetric source.
In most applications, the physical relevance of the Vaidya 
metric is limited to the
spacetime outside a star, with a different metric describing the star's
internal structure.
But black hole radiance \cite{swh} suggests use
of the Vaidya metric to model back-reaction effects 
for evaporating black holes \cite{hiscockII,balbinot} 
all the way upto the singularity.

The line element of the charged Vaidya solution is
\begin{equation}
ds^2 = - \vaidya du^2 - 2 \, du \, dr 
+ r^2 \lf d \theta ^2 + \sin ^2 \theta d \phi ^2 \rt \; .	\label{bv}
\end{equation}
The mass function $M(u)$ is the mass measured at future null infinity 
(the Bondi mass) and is in general a decreasing function of the outgoing null
coordinate, $u$. Similarly, the function $Q(u)$ describes the charge, measured
again at future null infinity. When $M(u)$ and $Q(u)$ are constant, the
metric reduces to the stationary Reissner-Nordstr\"om metric.
The corresponding stress tensor describes a purely electric Coulomb field,
\begin{equation}
F_{ru} = + {Q(u) \over r^2} \; ,
\end{equation}
and a null fluid with current
\begin{equation}
k_a = k \del _a u \; , \; \; k^2 = +{1 \over 4 \pi r^2} 
{\pl \over \pl u} \lf - M + {Q^2 \over 2r} \rt \; .
\end{equation}
In particular,
\begin{equation}
T_{uu} = {1 \over 8 \pi r^2} \left [\vaidya {Q^2(u) \over r^2} + 
{1 \over r}{\pl Q^2 (u) \over \pl u} - 2 {\pl M (u) \over \pl u} \right ] \; . 
\label{vstress}
\end{equation}

Like the Reissner-Nordstr\"om metric, the charged Vaidya metric is beset
by coordinate singularities. It is not known how to remove these 
spurious singularities for arbitrary mass and charge functions 
(for example, see \cite{senovilla}).
We shall simply choose functions for which the relevant integrations
can be done and continuation past the spurious singularities can be carried
out, expecting that the qualitative structure we find is robust.

Specifically, we choose the mass to be a decreasing linear function
of $u$, and the charge to be proportional to the mass:
\begin{equation}
M(u) \equiv au + b \equiv \tu \; , \; \; Q(u) \equiv \eta \tu \; , 
\label{linear}
\end{equation}
where $a < 0$ and $|\eta| \leq 1$, with $|\eta| = 1$ at extremality.
We always have $\tu \geq 0$.
With these choices, we can find an ingoing (advanced time) 
null coordinate, $v$, with which the line
element can be written in a ``double-null'' form:
\begin{equation}
ds^2 = - {g(\tu,r) \over a} d \tu \, dv + r^2 \lf d \theta ^2 + 
\sin ^2 \theta d \phi ^2 \rt \; .
\end{equation}
Thus
\begin{equation}
dv = {1 \over g(\tu,r)} \left [ \lf 1 - {2 \tu \over r} + {\eta^2 \tu ^2 
\over r^2} \rt {d \tu \over a} + 2 \, dr \right ] \; .	\label{g}
\end{equation}
The term in brackets is of the form $X(\tu,r) \, d \tu + Y(\tu,r) \, dr$. 
Since $X(\tu,r)$ and $Y(\tu,r)$ are 
both homogeneous functions, Euler's relation provides the integrating factor:
$g(\tu,r) = X(\tu,r) \tu + Y(\tu,r) r$.
Hence
\begin{equation}
{\pl v \over \pl r} = \f{r^2}{r^3 +{\tu \over 2a}(r^2 -2 \tu r+ \eta^2 \tu ^2)}
\label{dvdr}
\end{equation}
\begin{equation}
{\pl v \over \pl \tu} = \f{ {1 \over 2a}(r^2 - 2 \tu r + \eta^2 \tu ^2) }
{ r^3 + {\tu \over 2a} (r^2 - 2 \tu r + \eta^2 \tu ^2 ) } \; .
\end{equation}
{}From the sign of the constant term of the cubic,
we know that there is at least one positive zero. 
Then, calling the largest positive zero $r'$, we may factorize the cubic as
$(r - r')(r^2 + \beta r + \gamma)$. Hence
\begin{equation}
\gamma = - {\eta^2 \tu ^3 \over 2 a r'} > 0 \; , \; \;
\gamma - \beta r' = - {\tu ^2 \over 2 a} > 0 \; , \; \;
\beta - r' = {\tu \over 2a} < 0 \; .
\end{equation}
Consequently, the cubic
can have either three positive roots, with possibly a double root but not
a triple root, or one positive and two complex (conjugate) roots. 
We consider these in turn.

{\it i) Three positive roots}

When there are three distinct positive roots, 
the solution to Eq. (\ref{dvdr}) is
\begin{equation}
v = A \ln (r - r') + B \ln (r- r_2) + C \ln (r - r_1) \; ,
\end{equation}
where $r' > r_2 > r_1 > 0$, and 
\begin{equation}
A = {+ {r'}^2 \over (r'- r_2)(r' - r_1)} > 0 \; , \; \;
B = {- {r_2}^2 \over (r' - r_2)(r_2 - r_1)} < 0 \; , \; \;
C = {+ {r_1}^2 \over (r' - r_1)(r_2 - r_1)} > 0 \; .
\end{equation}
We can push through the $r'$ singularity by defining a new coordinate,
\begin{equation}
V_2 (v) \equiv e ^{v / A} = (r-r') (r - r_2) ^{B / A} (r - r_1) ^{C / A} \; ,
\end{equation}
which is regular for $r > r_2$. 
To extend the coordinates beyond $r_2$ we define
\begin{equation}
V_1 (v) \equiv k_2 + (- V_2) ^ {A / B} 
= k_2 + (r' - r)^{A / B} (r_2 - r) (r - r_1)^{C / B} \; ,
\end{equation}
where $k_2$ is some constant chosen to match $V_2$ and $V_1$ at some 
$r' > r > r_2$. $V_1(r)$ is now regular for $r_2 > r > r_1$. 
Finally, we define yet another coordinate,
\begin{equation}
V(v) \equiv k_1 + (-(V_1 - k_2))^{B / C} 
= k_1 + (r' - r )^{A / C} (r_2 - r) ^{B / C} (r - r_1) \; ,
\end{equation}
which is now free of coordinate singularities for $r < r_2$.
A similar procedure can be applied if the cubic has a double root.

{\it ii) One positive root}

When there is only one positive root, $v$ is singular only at $r = r'$:
\begin{equation}
v = A \ln (r - r') + {1 \over 2} B \ln (r^2 + \beta r + \gamma )
+ \f { 2 C -  B \beta}{\sqrt{4 \gamma - \beta ^2}} 
\arctan \lf \f{2r + \beta}{\sqrt{4 \gamma - \beta ^2}} \rt \; .
\end{equation}
We can eliminate this coordinate singularity by introducing a new coordinate
\begin{equation}
V(v) \equiv e^{v/A} = (r - r') (r^2 + \beta r + \gamma)^{B / 2A}
\exp \left [+ {2 C - B \beta \over A \sqrt{4 \gamma - \beta ^2}} 
\arctan \lf \f{2r + \beta}{\sqrt{4 \gamma - \beta ^2}} \rt \right ]  \; ,
\label{V1zero}
\end{equation}
which is well-behaved everywhere. The metric now reads
\begin{equation}
ds^2 = -g(\tu, r){A \over V(\tu,r)} \, {d\tu \over a} \, dV+r^2 d \Omega^2 \; .
\end{equation}
In all cases, to determine the causal structure 
of the curvature singularity we express
$dV$ in terms of $du$ with $r$ held constant. Now we note that,
since $\tu$ is the only dimensionful parameter, all derived dimensionful
constants such as $r'$ must be proportional to powers of $\tu$. For example,
when there is only positive zero, Eq. (\ref{V1zero}) yields
\begin{equation}
\left . dV \right | _r = d \tu \, {V \over \tu} \left [ {-r' \over r - r'} 
+ {B \over 2 A} {\beta r + 2 \gamma \over r^2 + \beta r + \gamma} 
+ {2C - B \beta \over A \sqrt{4 \gamma - \beta^2}} 
{1 \over 1 + \lf \f{2r + \beta} {\sqrt{4 \gamma - \beta^2}} \rt ^2}
{-2r \over {\sqrt{4 \gamma - \beta ^2}}} \right ] \; .
\end{equation}
Thus, as $r \rightarrow 0$, and using the fact that $A + B = 1$, we have
\begin{equation}
ds^2 \rightarrow - {Q^2(u) \over r^2} \, du^2 \; ,
\end{equation}
so that the curvature singularity is time-like.

\section{Patches of Spacetime}

Our working hypothesis is that the Vaidya spacetime, since it incorporates
radiation from the shrinking black hole, offers a more realistic background
than the static Reissner spacetime, where all back-reaction is ignored.
In this spirit, we can model the black hole's evolution by joining
patches of the collapse and post-evaporation (Minkowski) phases onto
the Vaidya geometry.

To ensure that adjacent patches of spacetime match along 
their common boundaries, we can calculate the stress-tensor at their 
(light-like) junction. The absence of a stress-tensor intrinsic to 
the boundary indicates a smooth match when there is no explicit source there.
Surface stress tensors are ordinarily computed by applying junction 
conditions relating discontinuities in the extrinsic curvature;
the appropriate conditions for light-like shells were obtained 
in \cite{barisrael}. However, we 
can avoid computing most of the extrinsic curvature tensors by using the
Vaidya metric to describe the geometry on both sides of a given boundary,
because the Reissner-Nordstr\"om and Minkowski spacetimes are
both special cases of the Vaidya solution.

Initially then, we have a collapsing charged spherically symmetric 
light-like shell. Inside the shell, region I, the metric must be 
that of flat Minkowski space; outside, region II, it must be the 
Reissner-Nordstr\"om metric, at least initially.
In fact, we can describe both regions together by a 
time-reversed charged Vaidya metric,
\begin{equation}
ds^2 = - \lf 1 - {2M(v) \over r} + {Q^2(v) \over r^2} \rt dv^2 +2 \, dv \, dr +
r^2 d \Omega ^2 \; ,
\end{equation}
where the mass and charge functions are step functions of the ingoing null
coordinate:
\begin{equation}
M(v) = M_0 \Theta ( v - v_0 ) \; , \; \; 
Q(v) = \eta M(v) \; .
\end{equation}
The surface stress tensor, $t^s_{vv}$, follows from Eq. (\ref{vstress}). Thus
\begin{equation}
t^s_{vv} = {1 \over 4 \pi r^2} \lf M_0 - {Q^2_0 \over 2r} \rt \; .
\end{equation}
The shell, being light-like, is constrained to move at 45 degrees on
a conformal diagram until it has collapsed completely. 
Inside the shell, the spacetime is guaranteed by Birkhoff's theorem to
remain flat until the shell hits $r = 0$, at which point a singularity forms.

Meanwhile, outside the shell, we must have 
the Reissner-Nordstr\"om metric. This is appropriate for all $r > r_+$. 
Once the shell nears $r_+$, however, one expects that quantum effects 
start to play a role.
For nonextremal ($|\eta| < 1$) shells, the Killing vector changes
character -- time-like to space-like -- as the apparent horizon is 
traversed, outside the shell. This permits a virtual pair, created by a vacuum
fluctuation just outside
or just inside the apparent horizon, to materialize by having one member
of the pair tunnel across the apparent horizon. Thus, Hawking radiation
begins, and charge and energy will stream out from the black hole.

We shall model this patch of spacetime, region III, by the Vaidya metric. 
This must be attached to the Reissner metric, region II, 
infinitesimally outside $r = r_+$. A smooth match requires that there be 
no surface stress tensor intrinsic to the boundary of the two regions.
The Reissner metric can be smoothly matched to 
the radiating solution along the $u =0$ boundary if $b = M_0$ 
in Eq. (\ref{linear}).

Now, using Eqs. (\ref{g}) and (\ref{V1zero}), 
one can write the Vaidya metric as
\begin{equation}
ds ^2 = - {g^2(\tu,r) A \over \vaidya V^2 } \, dV^2 + 
2 {g(\tu,r) A \over \vaidya V} \, dV \, dr \; .
\end{equation}
We shall assume for convenience that $g(r)$ has only one positive 
real root, which
we call $r'$. Then, since $V$ and $g$ both contain a factor $(r - r')$, Eq.
(\ref{V1zero}), the
above line element and the coordinates are 
both well-defined for $r > r_+ (\tu)$. 
In particular, $r = \infty$ is part of the Vaidya spacetime patch.
Moreover, the only solution with $ds^2 = dr = 0$ also
has $dV = 0$, so that there are no light-like marginally trapped
surfaces analogous to the Reissner $r_{\pm}$. In other words,
the Vaidya metric extends to future null infinity, ${\cal I}^+$, and hence
there is neither an event horizon, nor a second time-like singularity on
the right of the conformal diagram.

The singularity on the left exists until the radiation
stops, at which point one has to join the Vaidya solution to 
Minkowski space. This 
is easy: both spacetimes are at once encompassed by a Vaidya solution with 
mass and charge functions
\begin{equation}
M(u) = (au + b) \Theta (u_0 - u) \; , \; \; Q(u) = \eta M(u) \; .
\end{equation}
As before, the stress tensor intrinsic to the boundary at $u_0$ can be read off
Eq. (\ref{vstress}):
\begin{equation}
t^s_{uu} = {1 \over 4 \pi r^2} \left [ (au+b)-{(au+b)^2 \over 2r} \right ] \; ,
\end{equation}
which is zero if $u_0 = - b/a$, i.e., if $\tu =0$. This says simply that
the black hole must have evaporated completely before one can return to
flat space.

Collecting all the constraints from the preceding paragraphs, we can
put together a possible conformal diagram, as in Fig. 1. 
(We say ``possible'' because a similar analysis for an uncharged hole 
leads to a space-like singularity; thus our analysis demonstrates 
the possibility, but not the inevitability,
of the behavior displayed in Fig. 1.) 
Fig. 1 is a Penrose diagram showing the global structure of a spacetime
in which a charged imploding null shock wave collapses catastrophically 
to a point and subsequently evaporates completely. 
Here regions I and IV are flat Minkowski space,
region II is the stationary Reissner-Nordstr\"om spacetime, and region III
is our extended charged Vaidya solution. The zigzag line on the left 
represents the singularity, and the straight line separating region I
from regions II and III is the shell. The curve connecting the start of
the Hawking radiation to the end of the singularity is $r_+(\tu)$, which
can be thought of as a surface of pair creation. The part of region
III interior to this line might perhaps be better approximated by an
ingoing negative energy Vaidya metric.

{}From this cut-and-paste picture we see that, given some initial data set,
only regions I and II and part of region III can be
determined entirely; an outgoing ray starting at the bottom of the 
singularity marks the Cauchy horizon for these regions. 
Note also that there is no true horizon; the singularity is naked.
However, because the singularity is time-like,
Fig. 1 has the attractive feature that
predictability for the entire spacetime is restored 
if conditions at the singularity are known. It is tempting to
speculate that, with higher resolution, the time-like singularity 
might be resolvable into some dynamical Planck-scale object such as a D-brane.

\begin{figure} [!hbt]
\input{epsf}
\epsfxsize = 5.0in \epsfbox{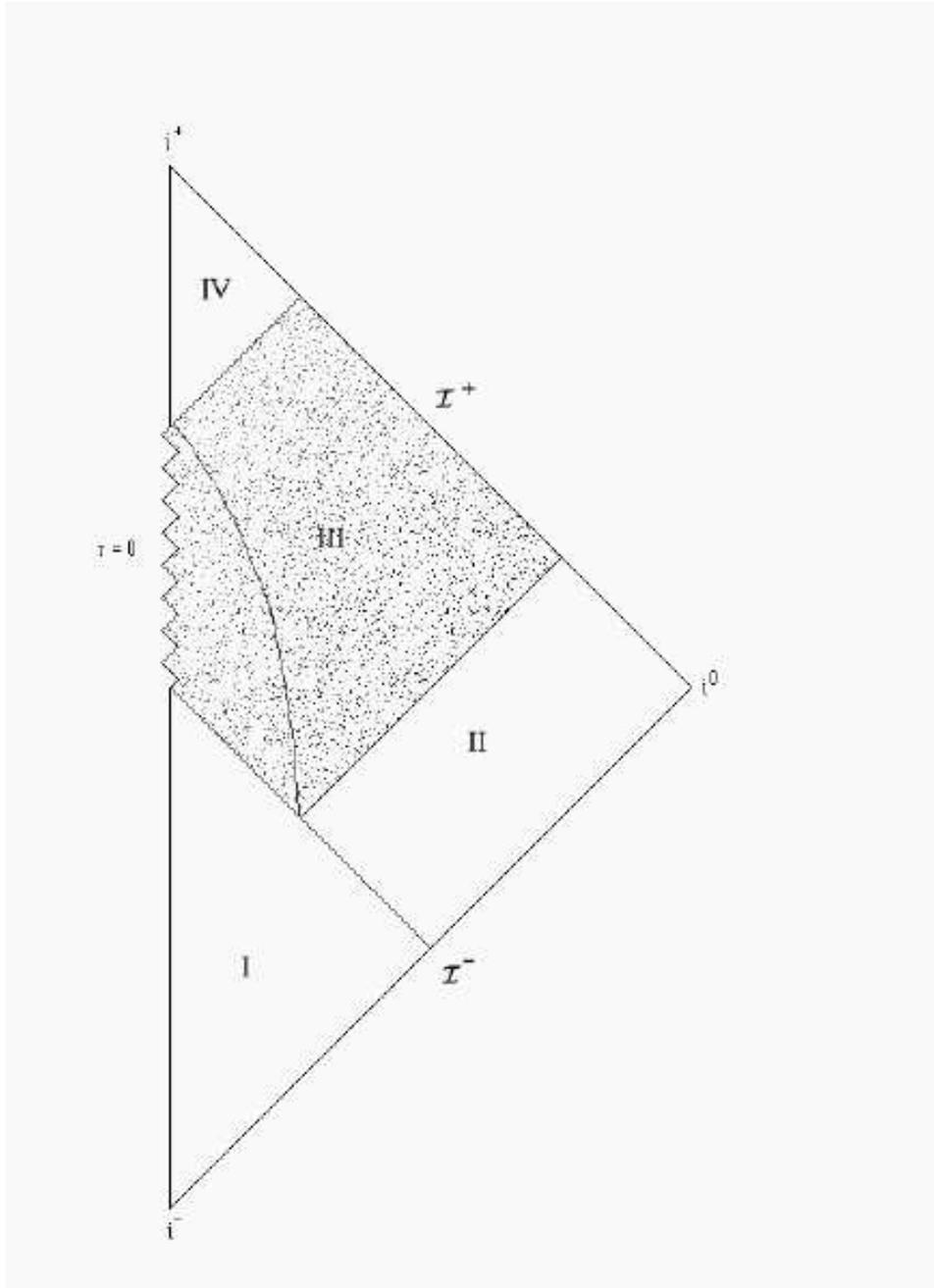}
\caption{Penrose diagram for the formation and evaporation of a charged black hole.}
\end{figure}